\documentclass[twocolumn,aps,pra,showpacs,superscriptaddress,nofootinbib,10pt]{revtex4-1}
\usepackage[latin1]{inputenc}
\usepackage{graphicx,dcolumn,bm}
\usepackage{subfigure}
\usepackage[hidelinks]{hyperref}
\usepackage{color}
\usepackage{amsfonts}
\usepackage{amssymb}
\usepackage{amsmath}
\usepackage{latexsym}
\usepackage{times,txfonts}
\usepackage{epstopdf}
\usepackage{braket}
\usepackage{float}

\usepackage{mathrsfs}
\usepackage{arydshln}
\usepackage[sans]{dsfont}

\usepackage{epsfig}

\definecolor{Blue}{rgb}{0.0,0.0,1}
\definecolor{Red}{rgb}{1,0.0,0.0}
\definecolor{Green}{rgb}{0,0.5,0.0}

\begin{document}

\title{ Enabling quantum non-Markovian dynamics by injection of classical colored noise}
\author{J. I. Costa-Filho}
\affiliation{Instituto de F\'{\i}sica de S\~{a}o Carlos, Universidade de S\~{a}o Paulo, CP 369, 13560-970, S\~{a}o Carlos, SP, Brazil}
\author{R. B. B. Lima}
\affiliation{Instituto de F\'{\i}sica de S\~{a}o Carlos, Universidade de S\~{a}o Paulo, CP 369, 13560-970, S\~{a}o Carlos, SP, Brazil}
\author{R. R. Paiva}
\affiliation{Instituto de F\'{\i}sica de S\~{a}o Carlos, Universidade de S\~{a}o Paulo, CP 369, 13560-970, S\~{a}o Carlos, SP, Brazil}
\author{P. M. Soares}
\affiliation{Instituto de F\'{\i}sica de S\~{a}o Carlos, Universidade de S\~{a}o Paulo, CP 369, 13560-970, S\~{a}o Carlos, SP, Brazil}
\author{W. A. M. Morgado}
\affiliation{Department of Physics, PUC-Rio and National Institute of Science and Technology for Complex Systems, Rua Marqu\^{e}s de S\~{a}o Vicente 225, 22453-900, Rio de Janeiro-RJ, Brazil}
\author{R. Lo Franco}
\affiliation{Dipartimento di Fisica e Chimica, Universit\`{a} di Palermo, via Archirafi 36, 90123 Palermo, Italy}
\affiliation{Dipartimento di Energia, Ingegneria dell'Informazione e Modelli Matematici, Universit\`{a} di Palermo, Viale delle Scienze, Edificio 9, 90128 Palermo, Italy}
\author{D. O. Soares-Pinto}
\affiliation{Instituto de F\'{\i}sica de S\~{a}o Carlos, Universidade de S\~{a}o Paulo, CP 369, 13560-970, S\~{a}o Carlos, SP, Brazil}

\begin{abstract}
The non-Markovian nature of quantum systems recently turned to be a key subject for investigations on open quantum system dynamics. Many studies, from its theoretical grounding to its usefulness as a resource for quantum information processing and experimental demonstrations, have been reported in the literature. Typically, in these studies, a structured reservoir is required to make non-Markovian dynamics to emerge. Here, we investigate the dynamics of a qubit interacting with a bosonic bath and under the injection of a classical stochastic colored noise. A canonical Lindblad-like master equation for the system is derived, using the stochastic wavefunction formalism. Then, the non-Markovianity of the evolution is witnessed using the Andersson, Cresser, Hall and Li measure. We evaluate the measure for three different noises and study the interplay between environment and noise pump necessary to generate quantum non-Markovianity, as well as the energy balance of the system. Finally, we discuss the possibility to experimentally implement the proposed model.
\end{abstract}


\maketitle

\graphicspath{{figures/}}


\section{Introduction}

The unavoidable interaction of every quantum system with its surroundings is the central topic of study in the theory of open quantum systems \cite{Breuer2007,Rivas2012,Carmichael199305,ALICKI,Davis197611,Weiss201203}. One of its main objectives is to understand how the system loses information to the environment and how it could be recovered \cite{Joos2003,Zurek2003}, which leads to the current interest in non-Markovian open quantum systems \cite{Rivas2014,Breuer2016,Wolf2008,Piilo2008,Vega2015,Modi2015,Brito2015}. Non-Markovian environments display the desired memory effects and information blackflows \cite{Rivas2014,Rivas2010,Bylicka2014,Addis2014,Rivas2010,Leggio2015} but, in turn, usually need to be highly structured \cite{Lloyd2001,Chin2012,Verstraete2009,man2015cavity,man2015harnessing,romero2012simple,lo2015switching,lo2016nonlocality,GonzlezGutirrez2016}.

The concept of non-Markovianity, although well understood in classical stochastic processes \cite{Kampen200705}, has no straightforward generalization to quantum systems. In classical probability theory, a stochastic process is Markovian if the conditional probability that it takes some value $x_n$ at the time $t_n$, given it had the value $x_{n-1}$ at time $t_{n-1}$, is independent of events prior to $t_{n-1}$ \cite{Rivas2014,Kampen200705}. In other words, the process is Markovian if the probability of going to some future state depends only on the present state and not on the previous ones, i.e., the process has no memory of its past states \cite{Rivas2014}. That definition does not work well in quantum mechanics, as we need to measure the system in its past states to formulate conditional probabilities. Since measurements in quantum mechanics disturb the system and, therefore, the conditional probabilities above, the definition of Markovianity would depend not only on the process to be analyzed but also on the choice of measurement scheme \cite{Rivas2014}, which is a undesirable drawback. 

In order to fix that, various definitions were proposed in the literature \cite{Rivas2012,ALICKI,Breuer2007,Accardi1985,Haikka2011,Addis2014,Neto2016}, but they are in most cases not equivalent to each other. On the one hand, the Rivas, Huelga, Plenio (RHP) definition \cite{Rivas2014} says that a quantum evolution is Markovian if it is CP-divisible, i.e, it satisfies a composition law analogous to the Chapman-Kolmogorov equation, which is obeyed by classical Markov processes \cite{Rivas2014,Kampen200705}. On the other hand, the Breuer, Lane, Piilo (BLP) definition \cite{Breuer2009} states that in Markovian processes the distinguishability of quantum states subject to the same evolution does not increase over time, which is a way to state that the process is memoryless. Both definitions are not equivalent, as are many other definitions in the literature, which shows that quantum non-Markovianity is in reality a multifaceted phenomenon.

As broad as the different definitions of quantum non-Markovianity is the plethora of its features and applications. Quantum non-Markovianity is related to preservation of coherence \cite{Viola1999}, energy backflows \cite{Guarnieri2016}, speedup of quantum speed limits \cite{Deffner2013}, violations of the Landauer bound \cite{Pezzutto2016}, formation of steady-state entanglement \cite{Huelga2012}, entanglement revivals \cite{lo2013dynamics,bellomo2007non,lo2014preserving,d2014recovering,aolita2015open} and is an obstacle to quantum Darwinism \cite{Galve2016}, for example. It has applications from quantum metrology \cite{Chin2012}, superdense coding \cite{Liu2016}, quantum cryptography \cite{Vasile2011} to quantum control \cite{Schmidt2011}. Recently, many experiments were conducted that verify or take advantage of non-Markovian features \cite{Bernardes2016,souza2013experimental,Orieux2015,Liu2011,Fanchini2014,Xu2013}. Finally, non-Markovianity is necessary for a realistic description of some quantum systems, such as strongly coupled systems \cite{Xiong2010,Batalhao2014}, some spin baths \cite{Krovi2007}, biological systems \cite{Chin2013}, complex nanostructures \cite{Crdenas2015} and  photosynthetic systems \cite{Chen2015}. As can be seen, quantum non-Markovianity is an invaluable resource for quantum technologies, which needs to be completely understood and harnessed.

In this paper, we show that one alternative to reservoir engineering is to induce quantum non-Markovianity by injection of classical noise \cite{Budini1999, Budini2000, Budini2001, James1998,Benedetti2013,Benedetti2014,lo2016overview}. A sufficiently strong noise could reverse the information flow from the system to the environment, thereby leading to memory effects. The procedure is as follows. We use the stochastic wave function's formalism \cite{Plenio1998,Dalibard1992}, where the state of the system is described by an ensemble of pure states, and the system density matrix is recovered by an averaging process. The environment is a thermal bath in the Born-Markov setting \cite{Breuer2007, Rivas2012}, while the classical noise is modeled by a stochastic Hamiltonian \cite{Budini1999, Budini2000, Budini2001}. Finally, the master equation of the system is derived using functional techniques \cite{NOVIKOV,moss1989noise,Vladimirov200208}, and a non-Markovianity measure \cite{Hall2014} is used to show that the evolution is indeed non-Markovian. We specifically use the Andersson, Cresser, Hall and Li (ACHL) \cite{Hall2014} measure, since it can be applied directly to the master equation of the system. The definitions of quantum non-Markovianity are in general nonequivalent, and so are its measures, thus we also relate the ACHL measure to other known measures of non-Markovianity. 

The structure of the paper is arranged as follows. In Sec.\ref{sec:II} we review the definition of quantum non-Markovianity, the ACHL non-Markovianity measure and its relation to other measures, while in Sec.\ref{sec:III} the master equation of the problem is derived. The results are discussed in Sec.\ref{sec:IV}, where different noises are applied to the master equation and its non-Markovianity is measured. An experimental proposal is made in Sec.\ref{sec:V} and, finally, Sec.\ref{sec:VI} contains the conclusion.


\section{Quantum non-Markovianity}
\label{sec:II}

\subsection{Definition}

In this paper we consider two quantum non-Markovianity definitions: the RHP \cite{Rivas2012} and BLP conditions \cite{Breuer2009}. The RHP condition states that a quantum process $\mathcal{E}(t,t_0)$ is Markovian if it is a CP-divisible map, i.e., a trace preserving, completely positive (CPTP) map such that, for any intermediate time, it can be broken into two CPTP maps. Namely, 
\begin{equation}
\label{eq:CP}
\mathcal{E}(t,t_0) = \mathcal{E}(t,t_1) \,\mathcal{E}(t_1,t_0), \ t_0 \leq t_1 \leq t,
\end{equation}
where $\mathcal{E}(t,t_1)$ and $\mathcal{E}(t_1,t_0)$ are CPTP maps. The BLP defines Markovianity as an evolution such that the trace distance between any two states decreases monotonically with time:
\begin{equation}
\frac{d}{dt}||\, \rho_1(t) - \rho_2(t) \,||_1 \leq 0,
\end{equation}
where $ \rho(t) = \mathcal{E}(t,t_0)\left [\rho \right] $ and  $|| X ||_1 = \mbox{tr} \sqrt{XX^{\dagger}}$. Physically, it means that, in a Markovian evolution, the indistinguishability between any two states cannot increase. In order to understand how the two definitions are related to each other, we need the concept of $k$-divisibility.

A quantum evolution $\mathcal{E}$ is positive if it takes positive operators (such as density operators) to positive operators, and is $k$-positive if $\mathds{1}_k \otimes \mathcal{E}$ is a positive evolution. If the evolution is $k$-positive for every $k \in \mathbb{N}$, then it is completely positive (CP). These concepts can be generalized to continuous in time evolutions: a $k$-divisible map \cite{Chruciski2014} is a $k$-positive map which can be arbitrarily broken into two other $k$-positive maps, and a CP-divisible map is simply a map which is $k$-divisible for every $k \in \mathbb{N}$. For simplicity, we call $1$-divisible maps as P-divisible maps. Now we are able to link the two definitions: the BLP and RHP conditions are equivalent to the map being P-divisible and CP-divisible, respectively, as shown in Ref. \cite{Chruciski2014}. Therefore, every non-Markovian evolution in the RHP sense is non-Markovian in the BLP sense, but not the converse.

\subsection{Decay rates measure}

The most general form of a completely positive, trace preserving Markovian (in the RHP sense) master equation is given by Lindblad's theorem \cite{Lindblad1976,ALICKI}: 
\begin{align}
\label{eq:LINDBLAD1}
&\frac{d \rho (t)}{dt} = \mathcal{L}_t\left[\rho(t)\right] = -i \left [H(t) \,,\, \rho (t)\right ] \nonumber \\ 
&+\sum_{\alpha, \,\beta}\gamma_{\alpha \beta} (t) \left [  A_\beta(t)\,\rho (t)\, A_\alpha^{\dagger}(t) - \frac{1}{2}\left \{ A_\alpha^{\dagger}(t)\,A_\beta(t) \,,\, \rho (t) \right \}  \right ],
\end{align}
where the $A_{\alpha}(t)$ are general operators acting on the system, $H(t)$ is a hermitian operator and $\gamma_{\alpha \beta} (t) \geq 0$, for every $\alpha$, $\beta$ and $t$. However, master equations obeying the form of Eq.(\ref{eq:LINDBLAD1}) and with possibly negative decay (or decoherence) rates $\gamma_{\alpha \beta} (t)$ can represent more general time-local master equations \cite{Hall2014}, such as non-Markovian ones. With that in mind, the negativity of the decay rates can be used as a measure of non-Markovianity, and is a very suitable measure, since it applies directly to the master equation. \par
The above measure, however, faces a problem: the Lindblad form is non-unique, and the same set of coefficients $\gamma_{\alpha \beta} (t)$ may generate different dynamics. To circunvent that issue, the canonical form \cite{Hall2014,Rivas2014} of Lindblad-like equations is used. This form is obtained by expressing the Lindbladian in a orthonormal operator space basis, i.e., a basis $\{ G_k \}^{d^2-1}_0$ such that $\mbox{tr} [G_i^{\dagger}G_j] = \delta_{ij}$, where $d$ is the dimension of the Hilbert space, and, for simplicity, $G_0 = \mathds{1}/\sqrt{d}$ \cite{Rivas2012}. In that basis,
\begin{align}
&A_\beta(t) = \sum_n a_{\beta n}(t)\,G_n, \ \ a_{\beta n}(t) = \mbox{tr}\left[G^{\dagger}_n V_{\beta}(t) \right ], \\
&A^{\dagger}_\alpha(t) = \sum_m a^*_{\alpha m}(t)\,G_m^{\dagger}, \ \ a^{*}_{\alpha m}(t) = \mbox{tr}\left [G_m V_{\alpha}^{\dagger}(t) \right ],
\end{align}
and the master equation is
\begin{align}
\label{eq:lindblad_g}
\frac{d \rho (t)}{dt} =& -i \left [H(t),\rho (t)\right ] \nonumber \\&+ \sum_{n,m}c_{nm} (t) \left [  G_n\,\rho (t) G_m^{\dagger} - \frac{1}{2}\left \{ G_m^{\dagger}G_n, \rho (t) \right \}  \right ],
\end{align}
where $c_{nm} = \sum_{\alpha,\,\beta}a_{\alpha m}^*(t)\,\gamma_{\alpha \beta}(t)\,a_{\beta n}(t)$. It is straightforward to show that the $c_{nm}$ form a Hermitian matrix $C$, just by using that the $\gamma_{\alpha \beta}(t)$ also form a Hermitian matrix. Since every Hermitian matrix can be diagonalized by a unitary operation, $C = UDU^{\dagger}$, being $U$ and $D$ unitary and diagonal matrices, respectively, we have that the coefficients of $C$ are given by $c_{nm} = \sum_{k}u_{nk}(t)\,\gamma_{k}(t)\,u^*_{mk}(t)$, where $u_{nk}(t)$ are coefficients of the unitary matrix $U$. Eq.(\ref{eq:lindblad_g}) is now written as
\begin{align}
& \frac{d \rho (t)}{dt} = -i \left [H(t) \,,\, \rho (t)\right ] \nonumber \\
&+ \sum_{k}\gamma_{k}(t)\left [  L_k(t)\,\rho (t)\,L_k^{\dagger}(t) - \frac{1}{2}\left \{ L_k^{\dagger}(t)\, L_k(t) \,,\, \rho (t) \right \}  \right ],
\label{eq:master_orto}
\end{align}
where
\begin{align}
&L_k(t) =  \sum_{n}u_{nk}(t) G_n,  \\
&L_k^{\dagger}(t) = \sum_{m}u^*_{mk}(t) G_m,
\end{align}
still form an orthonormal basis, since unitary operations preserve inner products.

The $\gamma_{k}(t)$ are the canonical decay rates and with them we can build the measures of non-Markovianity. Define \cite{Hall2014}
\begin{equation}
f(t) = \sum_{k=1}^{d^2-1} \mbox{max}\{ -\gamma_k(t),0 \} = \frac{1}{2} \sum_{k=1}^{d^2-1} \left [|\,\gamma_k(t)\,| - \gamma_k(t) \right]. \label{eq:ACHL_f}
\end{equation}
The decay rates or ACHL measure \cite{Hall2014} for the time interval $t_0 \leq \tau \leq t$ is
\begin{equation}
\label{eq:ACHL}
\mathcal{N}_{\textup{ACHL}} = \int_{t_0}^t f(\tau)\,d\tau.
\end{equation}
Note that $\mbox{max}\{ -\gamma_k(t),0 \}$ simply selects the negative part of each $\gamma_k(t)$, and is zero if it is strictly nonnegative. The term $f(t)$ sums the contributions of all negative decay rates, and $\mathcal{N}_{ACHL}$ is this contribution integrated along the time interval.

The ACHL measure is nonzero if at least one decay rate is negative, for any brief interval of time, since this is sufficient for the breakdown of CP-divisibility \cite{Rivas2014}. That condition, however, for most cases is not sufficient to break P-divisibility. Therefore, some BLP Markovian evolutions can be considered non-Markovian by this measure. 

\subsection{Relation with other measures}

The decay rates measure can be related to other non-Markovianity measures. The RHP measure \cite{Rivas2010} estimates the breakdown of CP-divisibility by computing how the intermediate dynamics deviates from complete positivity. Namely, given a quantum evolution $\mathcal{E}(t,t_0)$, it can be decomposed as in Eq.(\ref{eq:CP}), since the map is continuous in time. Supposing that, for some $t_1$, $\mathcal{E}^{-1}(t_1,t_0)$ exists, then the intermediate dynamical map
\begin{equation}
\mathcal{E}(t,t_1) = \mathcal{E}(t,t_0) \, \mathcal{E}^{-1}(t_1,t_0),
\end{equation}
is well defined. An evolution is non-Markovian if exists at least an intermediate time $t_1$ such that $\mathcal{E}(t,t_1)$ is not completely positive, and the latter is completely positive if its Choi matrix \cite{Choi1975,Jamiokowski1972}
\begin{equation} \label{eq:CHOI}
J \left (\mathcal{E}(t,t_1 ) \right ) = \frac{1}{d} \sum_{i,j=1}^{d} \ket{i} \bra{j} \otimes \mathcal{E} (t,t_1 ) \left[ \ket{i} \bra{j} \right],
\end{equation}
where $\ket{i}$ is an orthonormal basis for the system, is positive semidefinite. Since our evolution is trace preserving, Eq.(\ref{eq:CHOI}) is positive definite if its 1-norm is equal to unity \cite{Rivas2014}. With that in mind, we define
\begin{equation}
g(t) = \lim_{\epsilon \rightarrow 0^+}\frac{||J \left (\mathcal{E}(t+\epsilon,t ) \right ) ||_1 - 1}{\epsilon},
\end{equation}
and $g(t) > 0$ if and only if the evolution is non-Markovian. Then, the RHP measure is 
\begin{equation}
\mathcal{N}_{\textup{RHP}} = \int_{t_0}^{t} g(\tau)\,d\tau.
\end{equation}
The RHP measure is proportional to the ACHL measure \cite{Rivas2014,Rivas2010,Hall2014}: 
\begin{equation}
\mathcal{N}_{\textup{ACHL}} = \frac{d}{2}\mathcal{N}_{\textup{RHP}}.
\end{equation}
An interesting case is the qubit, where both measures are equivalent.

Another common measure is the BLP measure \cite{Breuer2009}, related to the BLP condition, and which measures the increase of distinguishability of quantum states. For any two states $\rho_1(t)$ and $\rho_2(t)$ undergoing the same evolution, their trace distance
\begin{equation}
D(\rho_1(t),\rho_2(t)) = \frac{1}{2} \mid \rho_1(t)-\rho_2(t) \mid\,,
\end{equation} 
is nonincreasing under completely positive evolutions \cite{NielsenChuang201206,Breuer2007},
\begin{equation}
\sigma(t,\rho_1,\rho_2) = \frac{d}{dt}D(\rho_1(t),\rho_2(t)) < 0.
\end{equation}

The BLP measure is defined as
\begin{equation}
\mathcal{N}_{\textup{BLP}} = \max_{\rho_1,\,\rho_2}\int _{\sigma >0}  \sigma(t',\rho_1,\rho_2) \,dt',
\end{equation}
where the integral is evaluated over time intervals on which $\sigma >0$. The measure is the maximum distance for two states for any possible initial states $\rho_1$ and $\rho_2$. Although its computation is non trivial, its relationship with the decay rates measure can be studied for the case of a single qubit \cite{Hall2014}.

Since qubits can be represented in Bloch form \cite{NielsenChuang201206},
\begin{equation}
\label{eq:BLOCH_FORM}
\rho = \frac{1}{2} \left (\mathds{1} + \vec{n} \cdot \vec{\sigma} \right ),
\end{equation}
where $\vec{n}$ is its Bloch vector, any master equation in Lindblad form,
\begin{equation}
\frac{d \rho (t)}{dt} = \mathcal{L}_t\left[\rho(t)\right],
\end{equation}
can be rewritten in terms of the Bloch vector as
\begin{equation}
\dot{\vec{n}} = D(t)\,\vec{n} + \vec{u}(t),
\end{equation}
where 
\begin{align}
D_{jk}(t) = \mbox{tr} \left[\sigma_j^{\dagger}\,\mathcal{L}_t(\sigma_k) \right ], \\
u_j(t) = \mbox{tr} \left[\sigma_j^{\dagger}\,\mathcal{L}_t(\mathds{1}) \right ],
\end{align}
are the matrix elements of the so called damping matrix $D(t)$ and the drift vector $\vec{u}(t)$, respectively \cite{Hall2014}.

The increase of trace distance between two arbitrary qubits is only possible if the increase occurs at infinitesimal distance. For any two infinitesimally separated qubits $\rho$ and $\rho + \delta \rho$, their squared trace distance is \cite{Hall2014}
\begin{equation}
D^2(\rho, \rho + \delta \rho) = \frac{1}{4} \delta \vec{n} \cdot \delta \vec{n},
\end{equation}
and it can be shown that it increases if the matrix $(D+D^T)(t)$ has a positive eigenvalue \cite{Hall2014}. For a qubit under amplitude damping \cite{Breuer2007,Rivas2012}, the case which will be studied in this article, this condition boils down to
\begin{equation} \label{eq:SOMAGAMMA}
\sum_{k=1}^{2} \gamma_k(t) < 0,
\end{equation} 
and a related measure is
\begin{align}
\label{eq:BLPqubit}
& h(t) = \mbox{max}\left \{ -\sum_{k=1}^{2}\gamma_k(t),0 \right \}, \\
& \mathcal{N}_{\textup{BLP}} = \int_{t_0}^{t} h(\tau)\,d\tau.
\end{align}

This condition is stronger than the required in the ACHL measure in Eq.(\ref{eq:ACHL_f}), since it needs the sum of all decay rates to be negative, and not just one of them. Then, for example, we could have Markovianity even in the presence of a negative decay rate, given the other decay rates were large enough to make the sum in Eq.(\ref{eq:BLPqubit}) positive. As a consequence, for single decoherence channels both measures are equivalent. 

The Bloch volume measure \cite{Lorenzo2013}, whose application is restricted to qubits, is another non-Markovianity measure. Since the volume, in the Bloch sphere, of accessible states under a CP quantum evolution only decreases, its increase can be related to non-Markovianity \cite{Rivas2014,Lorenzo2013}. Using the damping matrix,  it is possible to show \cite{Hall2014} that non-Markovianity emerges if and only if Eq.(\ref{eq:SOMAGAMMA}) holds, so there is an equivalence between the Bloch volume and BLP measures. Note that these measures are inequivalent in more general scenarios \cite{Rivas2014,Hall2014}.

In the next section we deduce the canonical master equation of a qubit in a bosonic bath and under the influence of a classical colored noise. After that, it is possible to use the ACHL measure and study the non-Markovian character of the evolution. 

\section{Stochastic pumping and amplitude damping}
\label{sec:III}

\subsection{System-environment evolution}

Our first step is deriving a master equation in Lindblad form for a system interacting with a bath. Let's consider an evolution described by the Hamiltonian
\begin{equation}
\label{eq:HALMILTON_ampdamp}
H = \frac{\omega_0}{2} \sigma_z + \sum_{r}\omega_r\, b_r^{\dagger}\,b_r + \sum_{r} g_r \left (\sigma_- \, b_r^{\dagger} + \sigma_+ \, b_r\right ),
\end{equation}
which describes a two-level system coupled to a bosonic bath \cite{Rivas2012}, and where $\omega_0$ is the system frequency, $b_r^{\dagger}$ ($b_r$) the creation (annihilation) operators of the environment and $g_k$ a coupling constant. The third term in the right hand side of Eq.(\ref{eq:HALMILTON_ampdamp}) is the Hamiltonian $H_{\textup{int}}$, responsible for the system-bath interaction, which can be written as
\begin{align}
H_{\textup{int}}& = \sigma_-\otimes\sum_{r} g_r \, b_r^{\dagger} + \sigma_+ \otimes \sum_{r} g_r \, b_r \nonumber \\ & \equiv A_1 \otimes B_1 + A_2 \otimes B_2.
\end{align}
The Liouville-von Neumann equation \cite{Ballentine199805} for the Hamiltonian in Eq.(\ref{eq:HALMILTON_ampdamp}), in the interaction picture, is
\begin{equation}
\frac{d}{dt}{\tilde{\rho}}(t) = -i \left [\tilde{H}_{\textup{int}}(t),\tilde{\rho} (t) \right] ,
\end{equation}
where $\tilde{A}$ means that the corresponding operator is in the interaction picture, and
\begin{align}
\tilde{H}_{\textup{int}}(t)& = e^{-i\omega_0 t}\sigma_-\otimes\sum_{r} g_r \, e^{-i \omega_r t} b_r^{\dagger} + e^{-i \omega_0 t} \sigma_+ \otimes \sum_{r} g_r \, e^{i \omega_r t} b_r \nonumber \\ 
& \equiv \tilde{A}_1(t) \otimes \tilde{B}_1(t) + \tilde{A}_2(t) \otimes \tilde{B}_2(t).
\label{eq:H_int_1}
\end{align}

In order to obtain the system's evolution, we must trace over the bath's degree of freedom \cite{Lindblad1976} and resort to the Born-Markov approximation \cite{Breuer2007,Rivas2012}. Namely, we assume that the reservoir correlation functions (RCFs) decay rapidly compared to the system evolution, so that we can work in a timescale where they are negligible and, therefore, memory effects are absent. The RCFs are
\begin{equation}
\mathcal{B}_{\alpha \beta} (t)= \mbox{tr}_B \left [ \tilde{B}_{\alpha}(t)B_{\beta} \,\rho_B \right],
\label{eq:CORRE_1}
\end{equation}
where $\alpha, \beta$ are the indexes related to the interaction term $H_{\textup{int}}$ and, in the case of Eq.(\ref{eq:H_int_1}), range from $1$ to $2$. The approximation is valid if we work in the weak coupling limit, i.e., we assume that the coupling constant $g_k$ is small, and the initial state is uncorrelated, 
\begin{equation}
\rho(0)=\rho_S(0) \otimes \rho_B,
\end{equation}
where $\rho_S$ is the system density operator and $\rho_B = \exp(- \beta H_S)/Z$, with $Z=\mbox{tr}[\exp(-\beta H_S)]$ and $\beta=1/T$, the environment density operator is in a thermal state. The RCFs of our problem are
\begin{align}
&\mathcal{B}_{11}(t) = \sum_{r}|g_r|^2e^{i\omega_r t}N(\omega_r) \\
&\mathcal{B}_{22}(t) = \sum_{r}|g_r|^2e^{-i\omega_r t}\left (1+N(\omega_r) \right ),
\end{align}
where  $N(\omega_r)=1/\left[ \exp \left( \omega_r / T \right) + 1 \right]$ is the density of states in the mode $\omega_r$ \cite{Breuer2007}, and the other RCFs $\mathcal{B}_{12}(t)$ and $\mathcal{B}_{21}(t)$ are zero. After performing the continuum limit,
\begin{equation}
\sum_{r} f(\omega_r) \rightarrow \int_{0}^{\infty}d\omega \frac{J(\omega)}{|g(\omega)|^2}f(\omega),
\end{equation}
where $J(\omega)$ is the spectral density of the bath and $f(\omega_k)$ is an arbitrary function of the bath frequencies, the (non-zero) RCFs become
\begin{align}
&\mathcal{B}_{11}(\tau) = \int_{0}^{\infty}d\omega \, J(\omega) \, e^{i\omega \tau}N(\omega), \\
&\mathcal{B}_{22}(\tau) = \int_{0}^{\infty}d\omega \, J(\omega) \, e^{-i\omega \tau}\left [1+N(\omega) \right ].
\end{align}
The decay rates are fourier transforms of the RCFs, and the final master equation for the system, in the Schr�dinger picture, where $\rho_S = \mbox{tr}_B(\rho)$, is \cite{Rivas2012}
\begin{align}
\label{eq:MASTER_BATH}
\frac{d }{dt} & \rho_S (t)  = \, -i\left [\frac{\omega_0}{2}  \sigma_z + H_{\textup{LS}}\,,\, \rho_S(t)\right] \nonumber \\
&+ \gamma\left( \omega_0 \right) N(\omega_0) \left [ \sigma_+\,\rho_S(t)\,\sigma_- - \frac{1}{2} \left \{\sigma_-\,\sigma_+ \,,\, \rho_S(t)\right \} \right ] \nonumber \\
&+ \gamma\left( \omega_0 \right) \left[1+N(\omega_0)\right]  \left [\sigma_- \,\rho_S(t)\,\sigma_+ - \frac{1}{2} \left \{\sigma_+ \,\sigma_-\,,\, \rho_S(t) \right \} \right ],
\end{align}
where $\gamma(\omega_0) = 2 \pi J(\omega_0)$ and $H_{\textup{LS}}$ is the Lamb shift Hamiltonian \cite{Rivas2012, Lidar2001},
\begin{equation}
 H_{LS} = \left (\mathcal{P} \int_{0}^{\infty} d\omega \, \frac{J( \omega ) \, [N( \omega )+1/2]}{\omega_0 - \omega}  \right ) \sigma_z
\end{equation}
where $\mathcal{P}$ represents the Cauchy principal value. The term $H_{LS}$ is simply a shift in the energy levels of the system.

\subsection{System-noise evolution}

The next step is to study the evolution of a quantum system under the influence of stochastic pumping of classical fields \cite{Budini2001,James1998}. In this model, the Liouville-von Neumann equation of the system is given by
\begin{equation}
\label{eq:MASTER_STOCH}
\frac{d}{dt}\rho(t) = -i \left [\frac{\omega_0}{2}\sigma_z + V(t),\rho(t)\right ].
\end{equation}
The term $V(t)$, responsible for the classical noise, is
\begin{align}
\label{eq:BARULHO}
V(t)& = i \left [ e^{-i\omega t} u(t)\sigma_- - e^{i\omega t} u^*(t)\sigma_+  \right] \nonumber \\ 
	& \equiv \xi_1(t)V_1 + \xi_2(t)V_2,
\end{align}
where $\xi(t)$ is a stochastic variable. We assume that it is a colored Gaussian noise with zero mean \cite{Kampen200705},
\begin{align}
&\overline{\xi(t)} = 0, \\
&\overline{\xi(t)\,\xi(t')} = \overline{\xi^*(t)\,\xi^*(t')} = 0, \\
&\overline{\xi^*(t)\,\xi(t')} = \overline{\xi(t)\,\xi^*(t')} \equiv \chi(t,t'),
\end{align} 
where the bar denotes an average over the stochastic realizations \cite{Budini2001}.

In order to deal with this stochastic behaviour we use the stochastic wave function formalism \cite{Carmichael199305} in which the state of an open quantum system is described by an ensemble of pure states $\rho(t) = \ket{\Psi (t)} \bra{\Psi (t)}$ \cite{Budini2001}. The density matrix of the system is recovered after an average over the stochastic realizations \cite{Kampen200705}  
\begin{equation}
\rho_S(t) = \overline{\rho(t)}.
\end{equation}
Taking the average over Eq.(\ref{eq:MASTER_STOCH}), we have that
\begin{equation}
\label{eq:MASTER_ENSEMBLE}
\frac{d}{dt}\rho_S (t) = i \left [\frac{\omega_0}{2}\sigma_z \,,\, \rho(t)\right ] -i \left [ \overline{V(t)\,\rho (t)}   -  \overline{ \rho (t) V(t)}   \right ].
\end{equation}
To calculate the terms $\overline{V(t)\,\rho (t)}$ and $\overline{ \rho (t) V(t)}$ we apply Novikov's theorem \cite{NOVIKOV,moss1989noise},
\begin{equation}
\label{eq:NOVIKOV2}
\overline{\xi(t)\,\rho[\xi]} = \int_{0}^{t}dt' \,\overline{\xi(t) \,\xi(t')}\, \overline{\frac{\delta \rho[\xi]}{\delta \xi(t')}},
\end{equation}
which considers the average of the product of a stochastic process $\xi(t)$ and its functional form $\rho[\xi]$, and where the last term in the right hand side is a functional derivative \cite{moss1989noise}. Following the steps of Refs. \cite{Budini2001,James1998} and assuming a weak coupling between system and pumping \cite{Budini1999}, Eq.(\ref{eq:MASTER_ENSEMBLE}) takes the form
\begin{align}
\label{eq:MASTER_NOVIKOV}
\frac{d}{dt}\rho_S (t) & =- i \left [\frac{\omega_0}{2}\sigma_z \,,\, \rho_S(t)\right] \nonumber \\ 
&- \sum_{\alpha,\,\beta} \int_{0}^{t}dt' \chi_{\alpha \beta}(t,t') \left[ V^\dagger_\alpha(t) ,  \left [V_\beta(t') \,,\, \rho_S (t) \right ] \right ].
\end{align}
Evaluating Eq.(\ref{eq:MASTER_NOVIKOV}), and assuming homogeneity in time correlations, i.e., $\chi_{\alpha\beta}(t,t') = \chi_{\alpha \beta}(t-t') \equiv \mathcal{S}(\tau)$, the master equation, in Lindblad form, is 
\begin{align}
\label{eq:MASTER_PUMP}
\frac{d }{dt} \rho_S (t) &=  -i\left [\frac{\omega_0}{2} \sigma_z+H_{\text{EFF}}(t) \,,\, \rho_S(t)\right] \nonumber \\ 
&+ \eta(t) \left [ \sigma_+ \,\rho_S(t) \,\sigma_- - \frac{1}{2} \left \{\sigma_- \,\sigma_+ \,,\, \rho_S(t)\right \} \right ] \nonumber \\ 
&+ \eta(t)  \left [\sigma_- \,\rho_S(t) \,\sigma_+ - \frac{1}{2} \left \{\sigma_+ \,\sigma_- \,,\, \rho_S(t) \right \} \right ],
\end{align}
where 
\begin{equation}
\eta (t) = 2 \int_{0}^{t}d\tau \, \mathcal{S}(\tau) \cos(\Delta \omega \, \tau),
\label{eq:eta_exp}
\end{equation}
and $H_{\text{EFF}}(t)$ appears due the coupling between the system and the stochastic pumping, and represents a shift in the system energy levels. It is defined as
\begin{equation}
 H_{\text{EFF}}(t) = \int_{0}^{t} d\tau \, \mathcal{S}(\tau) \sin(\Delta\omega \, \tau) \, \sigma_z, 
\end{equation}
and $\Delta \omega = \omega - \omega_0$. An analogous derivation, but using white Gaussian noise, is presented in Ref.\cite{Moussa1996}.

\subsection{Combined evolution}

Combining the evolutions of Eq.(\ref{eq:MASTER_BATH}) and Eq.(\ref{eq:MASTER_PUMP}), the complete master equation for the system takes the form
\begin{align}
\label{eq:FINAL_COMBINED}
\frac{d }{dt} \rho (t)  =&  \,-i\left [\frac{\omega_0}{2}  \sigma_z + H_{\textup{shift}}(t) \,,\, \rho(t)\right] \nonumber \\
&+ \gamma_1(t) \left [ \sigma_+ \,\rho(t) \,\sigma_- - \frac{1}{2} \left \{\sigma_- \,\sigma_+ \,,\, \rho(t)\right \} \right ] \nonumber \\ 
&+ \gamma_2(t)  \left [\sigma_- \,\rho(t) \,\sigma_+ - \frac{1}{2} \left \{\sigma_+ \,\sigma_- \,,\, \rho(t) \right \} \right ],
\end{align}
where the coefficients of the master equation
\begin{align}
&\gamma_1(t) =  \gamma N(\omega_0) + \eta(t) \label{eq:gamma1}, \\
&\gamma_2(t) =  \gamma \left[1+N(\omega_0) \right] + \eta(t), \label{eq:gamma2}
\end{align}
are the decay rates. The term $H_{\textup{shift}}(t)$ is the sum of the energy shifts $H_{\textup{LS}}$ and $H_{\text{EFF}}(t)$. Note that the master equation is already in canonical form, so the ACHL measure can be applied directly to its decay rates.

The decay rates in Eqs.(\ref{eq:gamma1}) and (\ref{eq:gamma2}) are composed of two terms, one related to the bath and the other to the noise pump. The first is always positive, and the second, which is time dependent, can be either positive or negative. Therefore, the non-Markovian character of the evolution, determined by the negativity of the decay rates \cite{Hall2014}, depends on the relative strength between system-bath and system-pump interactions. The evolution is Markovian when the temperature terms are greater, in absolute value, than the negative part of $\eta(t)$. Then, a question can be asked: which is the minimum temperature above which the evolution is Markovian or, equivalently, the maximum temperature under which the evolution shows non-Markovian effects? Following the ACHL criterion given in Eq.(\ref{eq:ACHL_f}) and (\ref{eq:ACHL}), it is determined by the point where at least one of the decay rates becomes negative. Since $\gamma_1(t)$ is always smaller than $\gamma_2(t)$, that point is when $ \gamma N(\omega_0) + \eta_{\textup{min}} = 0$, where 
\begin{equation}
\eta_{\text{min}} = \min_{t_0 \leq \tau \leq t} \eta(\tau).
\end{equation}
Then, a Markovianity temperature can be set as
\begin{equation}
T := \frac{ \omega_0}{\ln \left(1-\frac{\gamma}{\eta_{\text{min}}} \right ) }.
\end{equation}
This is the temperature below which non-Markovian features emerge.

The master equation for the qubit population $\rho_{11}(t)$ takes the simple form
\begin{equation}
\frac{d}{dt}\rho_{11}(t) = -\left[\gamma_1(t)+\gamma_2(t) \right ]\rho_{11}(t) + \gamma_1(t),
\end{equation}
and the other population is constrained by the unitarity of the trace of the density operator: $\rho_{22}(t) = 1 - \rho_{11}(t)$. It is simpler, however, to put the qubit density matrix in Bloch form, Eq.(\ref{eq:BLOCH_FORM}), and study the evolution of the $z$ component of the Bloch vector, $n_z(t) = \Braket{\sigma_z (t)}$:
\begin{equation}
\frac{d}{dt}n_z(t) = -\left[\gamma_1(t)+\gamma_2(t) \right ]n_z(t) + \gamma_1(t) - \gamma_2(t).
\end{equation}
In terms of $n_z(t)$, the average energy of the system is
\begin{equation}
\label{eq:ENERGIA}
\left \langle E\right \rangle = \mbox{tr} \left [\frac{\omega_0}{2}\sigma_z \, \rho(t) \right ] = -\frac{\omega_0}{2} \, n_z(t).
\end{equation}
For long times, $t \rightarrow \infty$, the system thermalizes and the rate of change of $n_z(t)$ is zero. Therefore, the average energy tends to the value
\begin{equation}
\left \langle E\right \rangle = \frac{\omega_0}{2} \left[ \frac{1}{2N(\omega_0) + 1 + 2\eta_{\infty}/\gamma} \right],
\end{equation}
where $\eta_{\infty}$ is the limit of $\eta(t)$ as $t \rightarrow \infty$.
Without the noise, the average energy converges to the value
\begin{equation}
\left \langle E_{\textup{no pump}} \right \rangle = -\frac{\omega_0}{2} \left[ \frac{1}{2N(\omega_0) + 1} \right],
\end{equation}
and therefore the energy difference $\left \langle E \right \rangle-\left \langle E_{\textup{no pump}} \right \rangle$ is
\begin{equation}
\Delta E = \frac{\omega_0}{2} \frac{\gamma}{2 \eta_{\infty}}\left[ 1 + \left (\frac{2 \eta_{\infty}}{\gamma \left (2N(\omega_0)+1 \right )} \right ) \right]^{-1}.
\label{eq:delta_e}
\end{equation}
Note that, since $\eta_{\infty}$ is usually very small, if we vary $N(\omega_0)$ the energy difference will not change considerably. 


\section{Role of stochastic noise}
\label{sec:IV}

\subsection{Exponential noise}

In this section we study Eq.(\ref{eq:FINAL_COMBINED}) for different classical stochastic noise injections, which are defined by their correlation functions $\mathcal{S}(\tau)$. The first noise correlation which we analyze is an exponential decay \cite{Benedetti2014} (which characterises the Ornstein-Uhlenbeck process),
\begin{equation}
\mathcal{S}_{\textup{OU}}(\tau) = \frac{\Omega}{2 \tau_c}e^{-\tau/\tau_c},
\end{equation}
where $\Omega$ is an amplitude and $\tau_c$ the correlation time. The $\Omega$ must remain small, since otherwise the weak coupling assumption would be violated. The exact formula for $\eta(t)$ is
\begin{equation}
\label{eq:ETA}
\eta (t) = \frac{\Omega}{1 + (\Delta \omega \, \tau_c)^2} \left [ 1 + \sqrt{1 + (\Delta \omega \, \tau_c)^2} \,e^{-t/\tau_c}\sin(\Delta \omega \, t-\phi) \right ],
\end{equation}
where
\begin{equation}
\phi = \sin^{-1}\left [ \frac{1}{\sqrt{1 + (\Delta \omega \, \tau_c)^2}}\right].
\end{equation}
Note that the function $\eta(t)$ is composed of a constant term and a damped oscillating term. 

\begin{figure}[!h]	
	\centering 
	\includegraphics[width=7cm]{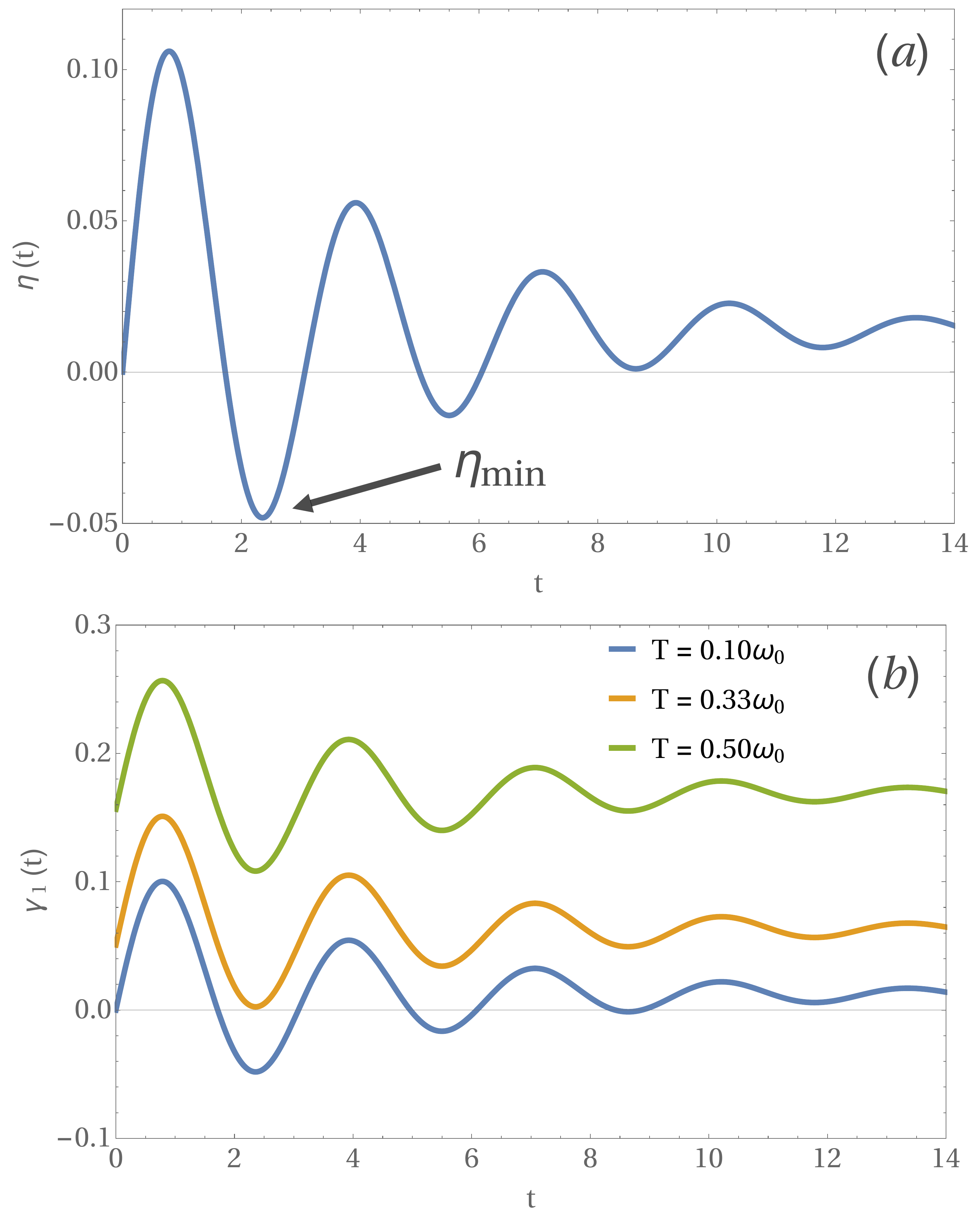}
	\caption{(Color Online) (a) $\eta(t)$, defined in Eq.(\ref{eq:eta_exp}), and (b) $\gamma_1(t)$, defined in Eq.(\ref{eq:gamma1}), as a function of time using the exponencial pump for three different temperatures: $T=0.10 \,\omega_0$ (blue), $T=0.33 \,\omega_0$ (yellow) which characterize the limit to find Markovian dynamics in the system  and $T=0.50 \,\omega_0$ (green). All the curves used: $\Delta\omega=2$, $\tau_c=4.25$, $\Omega=0.91$, $\gamma=1$.}
	\label{fig:ETAGAMMA26}
\end{figure}

We will consider a system where $T=c \, \omega_0$ wherein $c$ is a constant, i.e., 
We define the temperature as a function of the frequency of the system. Fixing all other parameters but the product $\Delta \omega \, \tau_c$, $\eta (t)$ reaches a global minimum value when $\Delta \omega \, \tau_c=8.5$. With these conditions, the only remaining free parameter is $\Omega$. Now, we want to find the Markovian to non-Markovian transition, i.e., the smallest value of $\Omega$ for which the decay rates have a negative part. Physically, we are looking for the smallest pump intensity for which the system reaches the Markovianity limit, where this limit identifies the border between the two regimes. That value is $\Omega = 0.91$, which corresponds to the situation when the decay rate $\gamma_1(t)$ touches the horizontal axis. The coefficients $\eta(t)$ and $\gamma_1(t)$ are plotted in Fig. \ref{fig:ETAGAMMA26}.

\begin{figure}[!h]	
	\centering 
	\includegraphics[width=7cm]{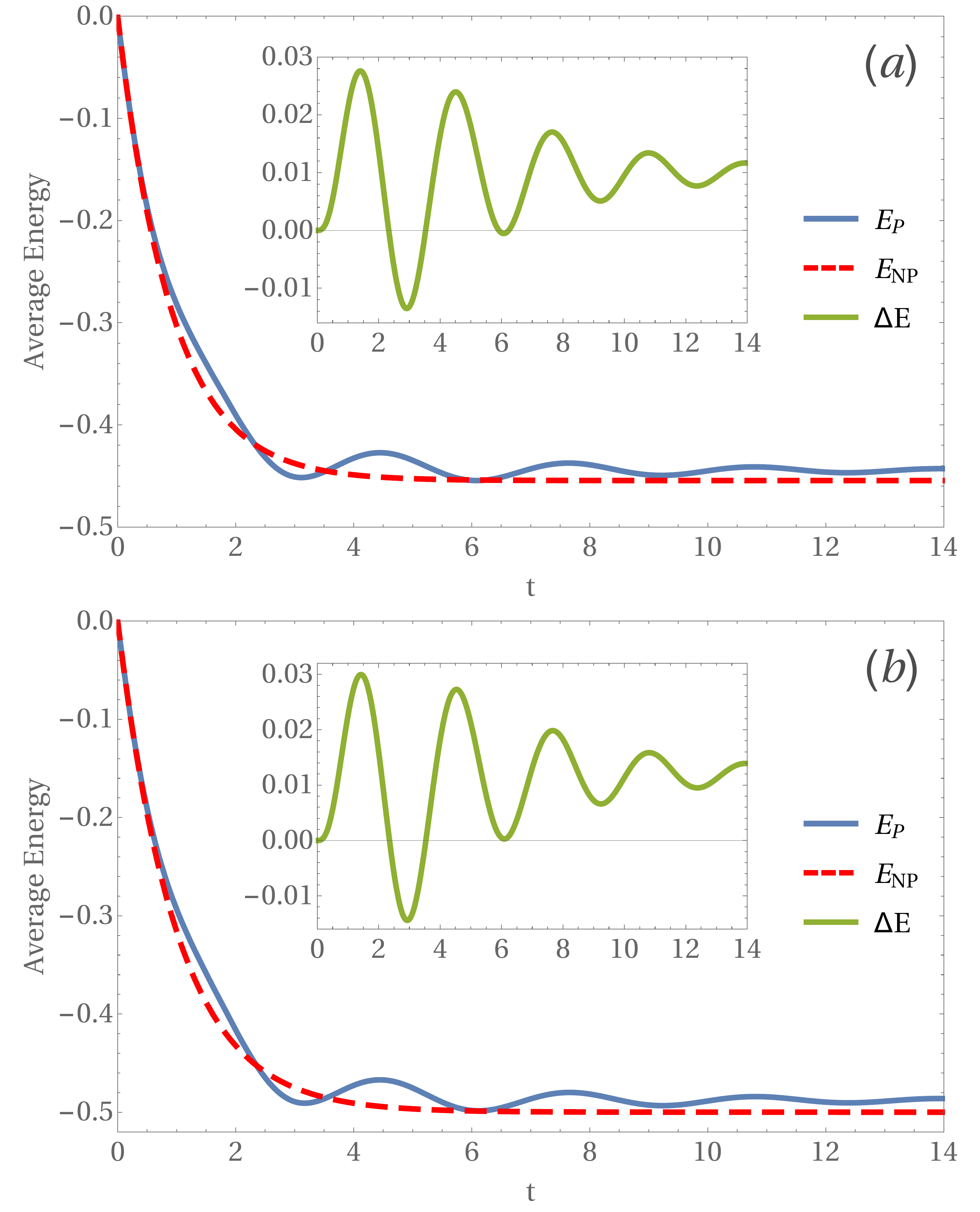}
	\caption{(Color online) Average energy of the system, in units of $\omega_0$, with ($E_P$) and without ($E_{NP}$) exponencial pump, and their difference ($\Delta E=E_P-E_{NP}$), in (a) Markovian regime, with $T=0.33 \,\omega_0$, and (b) non-Markovian regime, with $T=0.10 \,\omega_0$. The parameters used were: $\Delta\omega=2$, $\tau_c=4.25$, $\Omega=0.91$, $\gamma=1$. We can observe that the energy difference in the two regimes are equivalent, thus, the energy cost to generate non-Markovianity is low.}
	\label{fig:ENERGYOU}
\end{figure}

If we assume that the initial state of the system is $\rho(0) = | + \rangle \langle + |$, and $\Ket{+}=\frac{1}{\sqrt{2}}\left( \Ket{0} + \Ket{1} \right)$ (which implies that $n_z(0)=0$), then the average energy of the system can be calculated. The average energy with ($E_P$) and without ($E_{NP}$) pump, as well as the difference ($\Delta E = E_P-E_{NP}$), at $T = 0.1\omega_0$, are shown in Fig. \ref{fig:ENERGYOU}(a), in units of $\omega_0$. Note that the average energy tends to a bigger value when the pump acts on the system, than when it is absent.

Note that for short times, i.e., $t \ll \tau_c$, we have
\begin{equation}
\eta_{\textup{OU}}(t) \simeq \frac{\Omega t}{\tau_c}.
\end{equation}
For the opposite limit, $t \rightarrow \infty$,
\begin{equation}
\eta_{\textup{OU}}\rightarrow \frac{\Omega}{1 + (\Delta \omega \tau_c)^2},
\end{equation}
and the average energy tends to 
\begin{equation}
\Braket{E} \rightarrow  \frac{\omega_0}{2} \left[ \frac{1}{2N(\omega_0) + 1 + \frac{2 \Omega}{1+\left( \Delta\omega\,\tau_c \right)^2}} \right].
\end{equation}

\begin{figure}[!h]	
	\centering 
	\includegraphics[width=7cm]{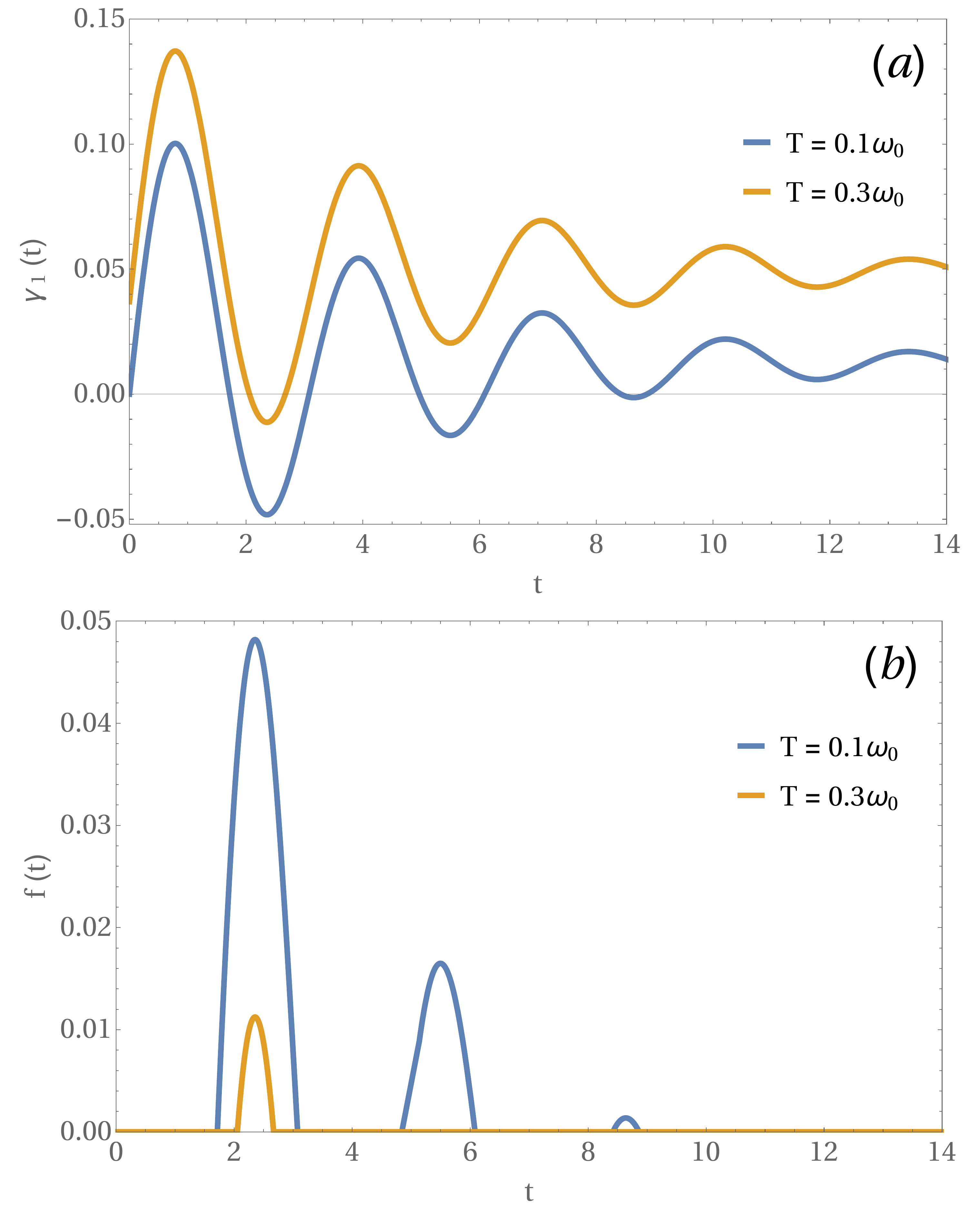}
	\caption{(Color online) (a) Decay rate $\gamma_1(t)$, Eq.(\ref{eq:gamma1}), and (b) ACHL measure $f(t)$, Eq.(\ref{eq:ACHL_f}), as a function of time in non-Markovian regime using the exponencial pump for two different temperatures, $T=0.1\,\omega_0$ (blue) and $T=0.3\,\omega_0$ (yellow). (b) The value obtained for the measure $\mathcal{N}_{\textup{ACHL}}=0.0546$ for $T=0.1\,\omega_0$ and $\mathcal{N}_{\textup{ACHL}}=0.0046$ for $T=0.3\,\omega_0$. The parameters used were: $\Delta\omega=2$, $\tau_c=4.25$, $\Omega=0.91$, $\gamma=1$.}
	\label{fig:GAMMAACHL15}
\end{figure}

Now the temperature is lowered to $T=0.1 \,\omega_0$, where non-Markovian effects are present. The average energy is plotted in Fig. \ref{fig:ENERGYOU}(b), and the decay rate $\gamma_1(t)$ and the ACHL measure $f(t)$ are plotted in Fig.\ref{fig:GAMMAACHL15}. We can see that the average energy increased compared to the Markovian case, due to its dependence with the density of the states, $\Braket{E} \propto 1/N(\omega_0)$. For the decay rate $\gamma_1(t)$, negative regions can be observed, which is a sign of non-Markovianity. These regions generate the peaks in the ACHL measure. The values obtained for the measure were $\mathcal{N}_{\textup{ACHL}}=0.0546$ for $T=0.1 \,\omega_0$ and $\mathcal{N}_{\textup{ACHL}}=0.0046$ for $T=0.3 \,\omega_0$.

\subsection{Squared exponential noise}

\begin{figure}[!h]	
	\centering 
	\includegraphics[width=7cm]{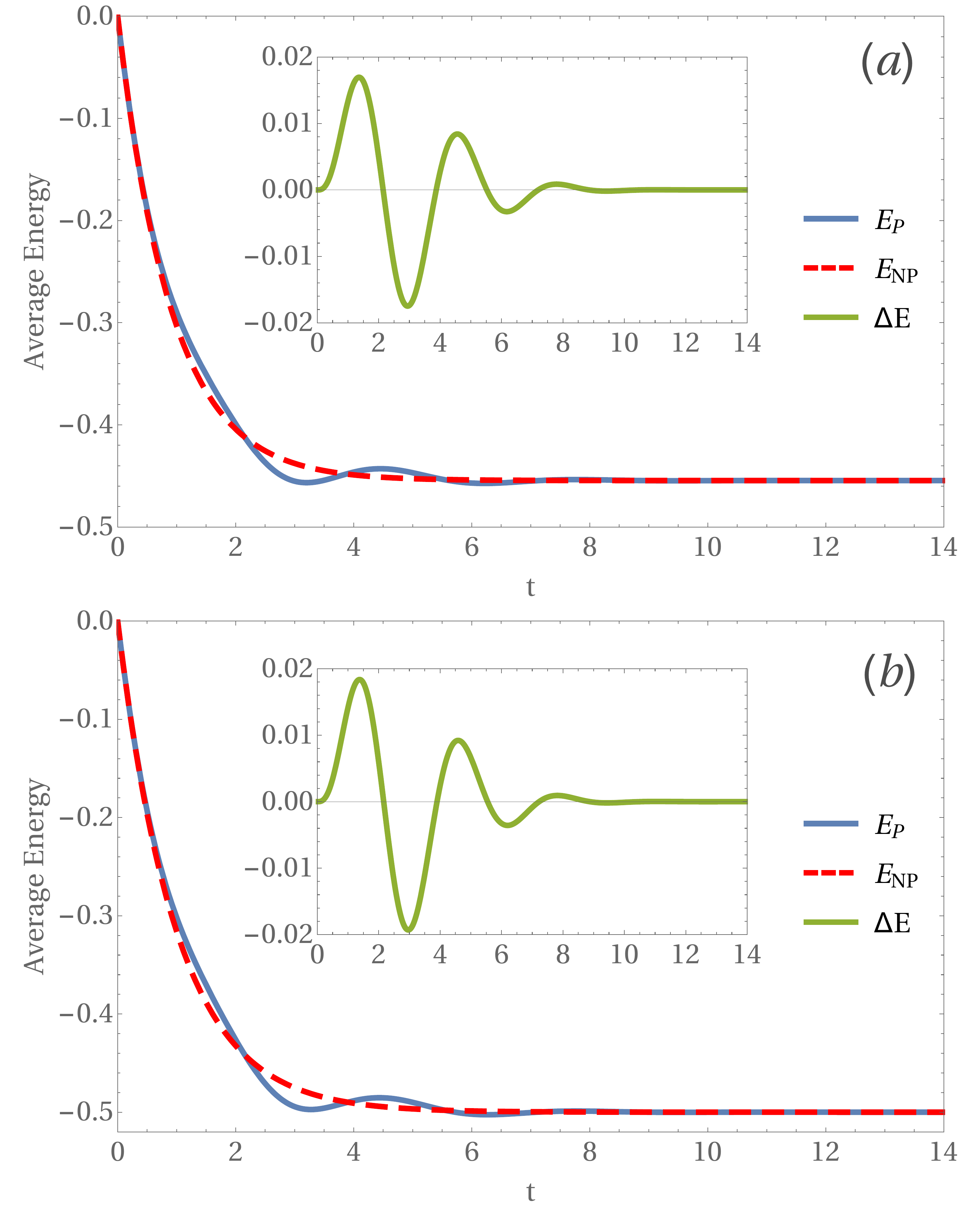}
	\caption{(Color online) Average energy of the system with ($E_P$) and without ($E_{NP}$) squared exponencial pump, and their difference ($\Delta E=E_P-E_{NP}$), in (a) Markovian regime, with $T=0.33\,\omega_0$, and (b) non-Markovian regime, with $T=0.10\,\omega_0$. The parameters used were: $\Delta\omega=2$, $\tau_c=4.25$, $\Omega=0.47$, $\gamma=1$.}
	\label{fig:ENERGYES}
\end{figure}

\begin{figure}[!h]	
	\centering 
	\includegraphics[width=7cm]{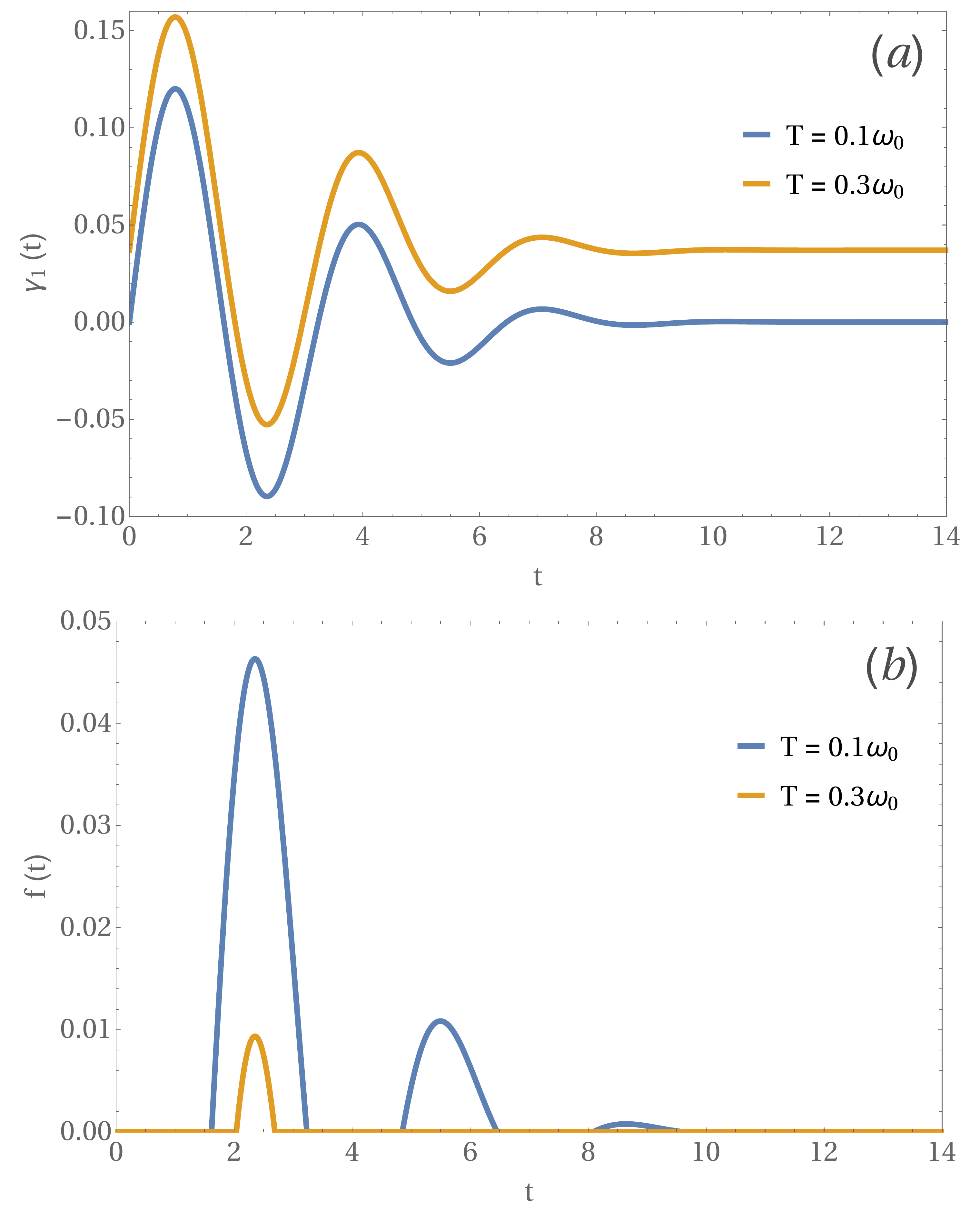}
	\caption{(Color online) (a) Decay rate $\gamma_1(t)$, Eq.(\ref{eq:gamma1}), and (b) ACHL measure $f(t)$, Eq.(\ref{eq:ACHL_f}), as a function of time in non-Markovian regime using the squared exponencial pump for two different temperatures, $T=0.1\,\omega_0$ (blue) and $T=0.3\,\omega_0$ (yellow). (b) The value obtained for the measure were $\mathcal{N}_{\textup{ACHL}}=0.0585$ for $T=0.1\,\omega_0$ and $\mathcal{N}_{\textup{ACHL}}=0.0040$ for $T=0.3\,\omega_0$. The parameters used were: $\Delta\omega=2$, $\tau_c=4.25$, $\Omega=0.47$, $\gamma=1$.}
	\label{fig:GAMMAACHL15ES}
\end{figure}

The squared exponential correlation function is \cite{Benedetti2014},
\begin{equation}
\mathcal{S}_{\textup{SE}}(\tau) = \frac{\Omega}{\sqrt{\pi}\tau_c}e^{-(t/\tau_c)^2},
\end{equation}
and for the same parameters as in the exponential case, we reach the Markovian limit with $\Omega = 0.47$. The average energies for $T=0.33 \,\omega_0$ and $T=0.10 \,\omega_0$ are plotted in Fig. \ref{fig:ENERGYES}. Note that, for this noise, the difference in the energies is almost negligible. For short times,
\begin{equation}
\eta_{\textup{SE}}(t)\simeq \frac{2 \Omega t}{\sqrt{\pi}\tau_c},
\end{equation}
and for the long time limit,
\begin{equation}
\eta_{\textup{SE}}\rightarrow \Omega \,e^{- \left ( \Delta \omega \tau_c / 2\right )^2}.
\end{equation}
The average energy tends to
\begin{align}
\Braket{E} \rightarrow & \,\frac{\omega_0}{2} \left[ \frac{1}{2N(\omega_0) + 1 + 2 \Omega \,e^{-\left( \Delta\omega \, \tau_c /2\right)^2}} \right].
\end{align}
The decay rate $\gamma_1(t)$ and the ACHL measure $f(t)$ are plotted in Fig. \ref{fig:GAMMAACHL15ES}. In this case, The values for the measure were $\mathcal{N}_{\textup{ACHL}}=0.0585$ for $T=0.10 \,\omega_0$ and  $\mathcal{N}_{\textup{ACHL}}=0.0040$ for $T=0.30 \,\omega_0$.

\subsection{Power law noise}

\begin{figure}[!h]	
	\centering 
	\includegraphics[width=7cm]{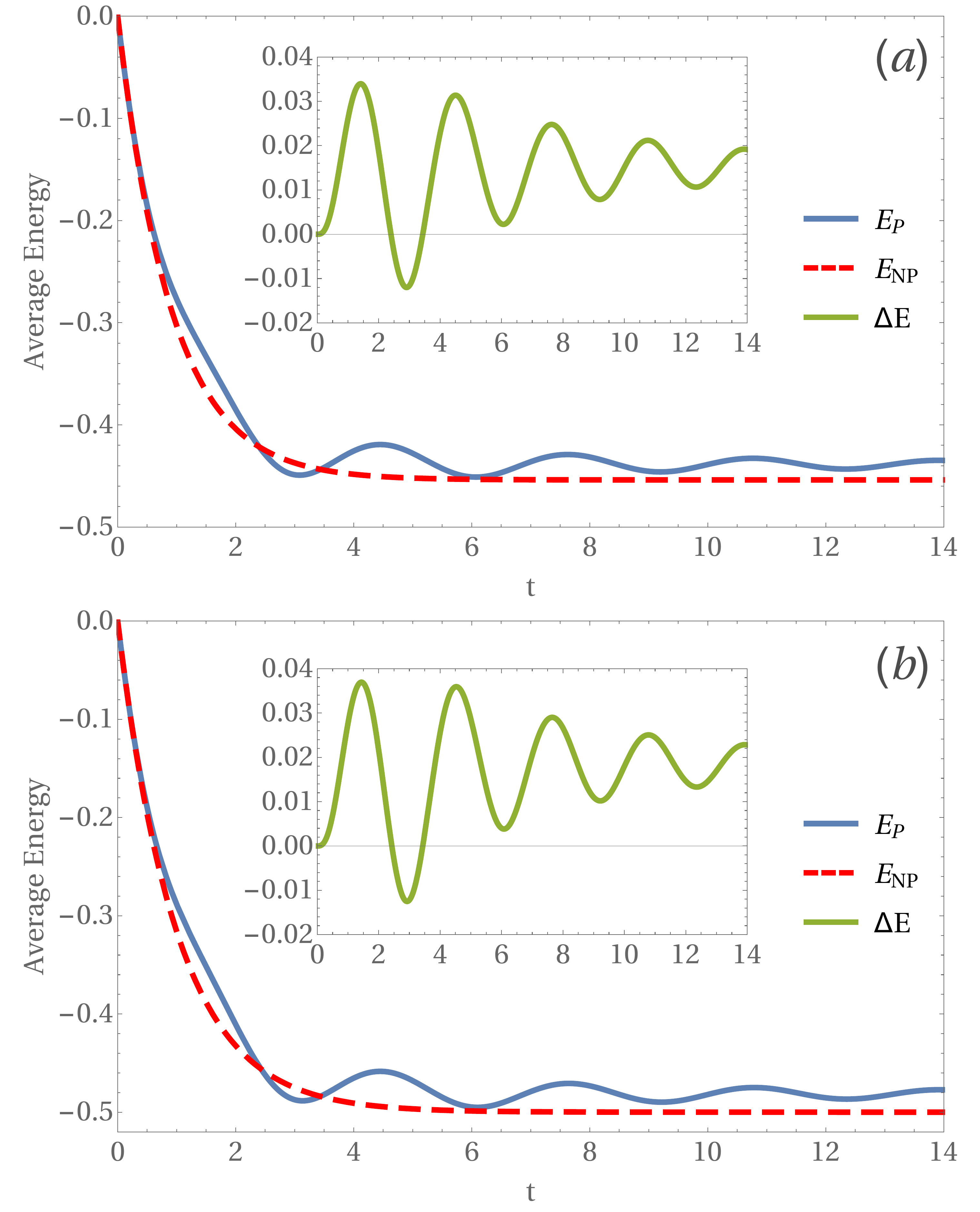}
	\caption{(Color online) Average energy of the system with ($E_P$) and without ($E_{NP}$) power law pump, and their difference ($\Delta E=E_P-E_{NP}$), in (a) Markovian regime, with $T=0.33\,\omega_0$, and (b) non-Markovian regime, with $T=0.10\,\omega_0$. The parameters used were: $\Delta\omega=2$, $\tau_c=10$, $\Omega=1.35$, $\gamma=1$.}
	\label{fig:ENERGYPL}
\end{figure}

\begin{figure}[!h]	
	\centering 
	\includegraphics[width=7cm]{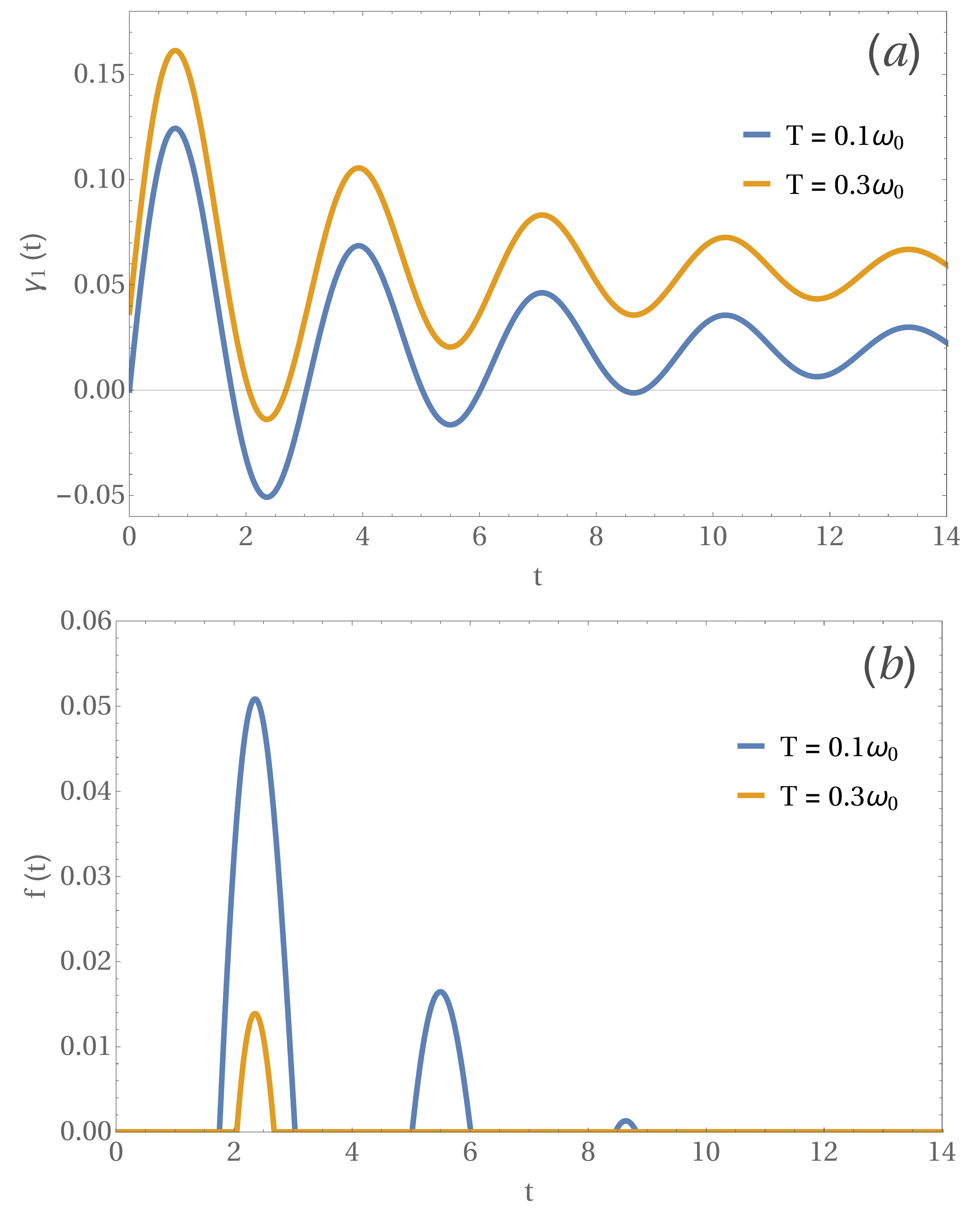}
	\caption{(Color online) (a) Decay rate $\gamma_1(t)$, Eq.(\ref{eq:gamma1}), and b) ACHL measure $f(t)$, Eq.(\ref{eq:ACHL_f}), as a function of time in non-markovian regime using the power law pump for two different temperatures, $T=0.1\,\omega_0$ (blue) and $T=0.3\,\omega_0$ (yellow). (b) The value obtained for the measure were $\mathcal{N}_{\textup{ACHL}}=0.0532$ for $T=0.1\,\omega_0$ and $\mathcal{N}_{\textup{ACHL}}=0.0058$ for $T=0.3\,\omega_0$. The parameters used were: $\Delta\omega=2$, $\tau_c=10$, $\Omega=1.35$, $\gamma=1$.}
	\label{fig:GAMMAACHL15PL}
\end{figure}

For the power law noise \cite{Benedetti2014},
\begin{equation}
\mathcal{S}_{\textup{PL}}(\tau) = \frac{\left( \alpha-1 \right)}{2}\frac{\Omega}{\tau_c}\frac{1}{\left(\tau / \tau_c +1 \right )^{\alpha}},
\end{equation}
where $\alpha > 2$, we have a global minimum (using $\alpha = 3$) for $\Delta\omega \, \tau_c=20$, and the Markovian limit is reached with $\Omega = 1.35$. The average energies for $T=0.33\,\omega_0$ (Markovian regime) and $T=0.10\,\omega_0$ (non-Markovian regime) are plotted in Fig. \ref{fig:ENERGYPL}. The short time limit is
\begin{equation}
\eta_{\textup{PL}}(t) \simeq \frac{2 \Omega t}{\tau_c},
\end{equation}
For $t \rightarrow \infty$, we were not able to find an analytical expression, but the decay rate $\gamma_1(t)$ and the ACHL measure $f(t)$ are plotted in Fig. \ref{fig:GAMMAACHL15PL}(b). The value obtained for the measure were $\mathcal{N}_{\textup{ACHL}}$ is $0.0532$ for $T=0.1\,\omega_0$ and $\mathcal{N}_{\textup{ACHL}}$ is $0.0058$ for $T=0.3\,\omega_0$.

\subsection{Comparisons} \label{MEDIDAS}

\begin{figure}[!h]	
	\centering 
	\includegraphics[width=7cm]{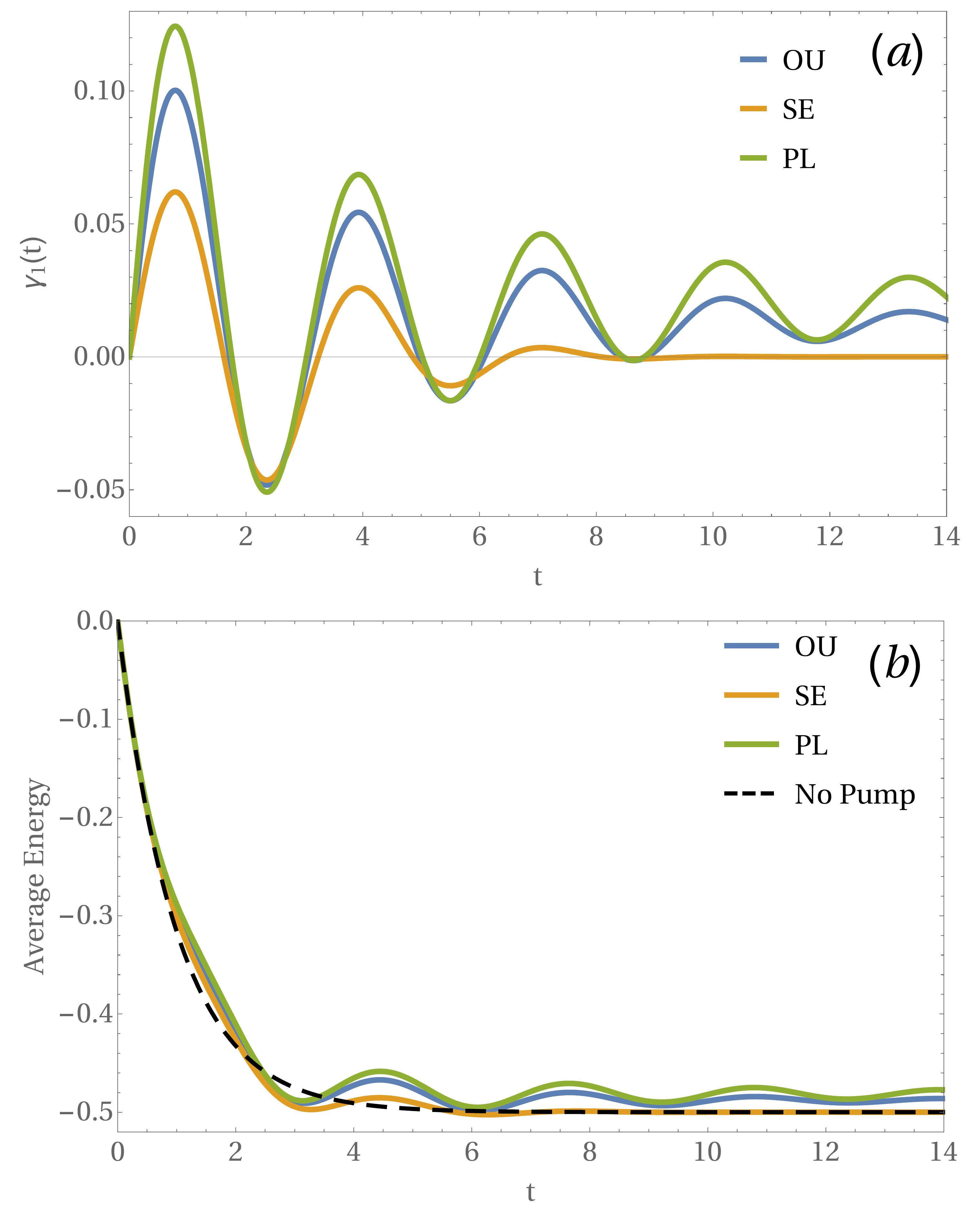}
	\caption{(Color online) (a) Decay rate $\gamma_1(t)$, Eq.(\ref{eq:gamma1}), as a function of time in non-Markovian regime for different noises. The parameters used were: $\Delta\omega=2$, $\tau_c=4.25$, $\Omega=0.91$ (OU); $\Delta\omega=2$, $\tau_c=4.25$, $\Omega=0.47$ (SE); $\Delta\omega=2$, $\tau_c=10$, $\Omega=1.35$ (PL). (b) Average energy for different noises. The parameters used were: $\Delta\omega=2$, $\tau_c=4.25$, $\Omega=0.91$ (OU); $\Delta\omega=2$, $\tau_c=4.25$, $\Omega=0.47$ (SE); $\Delta\omega=2$, $\tau_c=10$, $\Omega=1.35$ (PL). All the curves used: $\gamma=1$, $T=0.1\,\omega_0$.}
	\label{fig:GAMMA15ALL}
\end{figure}

\begin{figure}[!h]	
	\centering 
	\includegraphics[width=7cm]{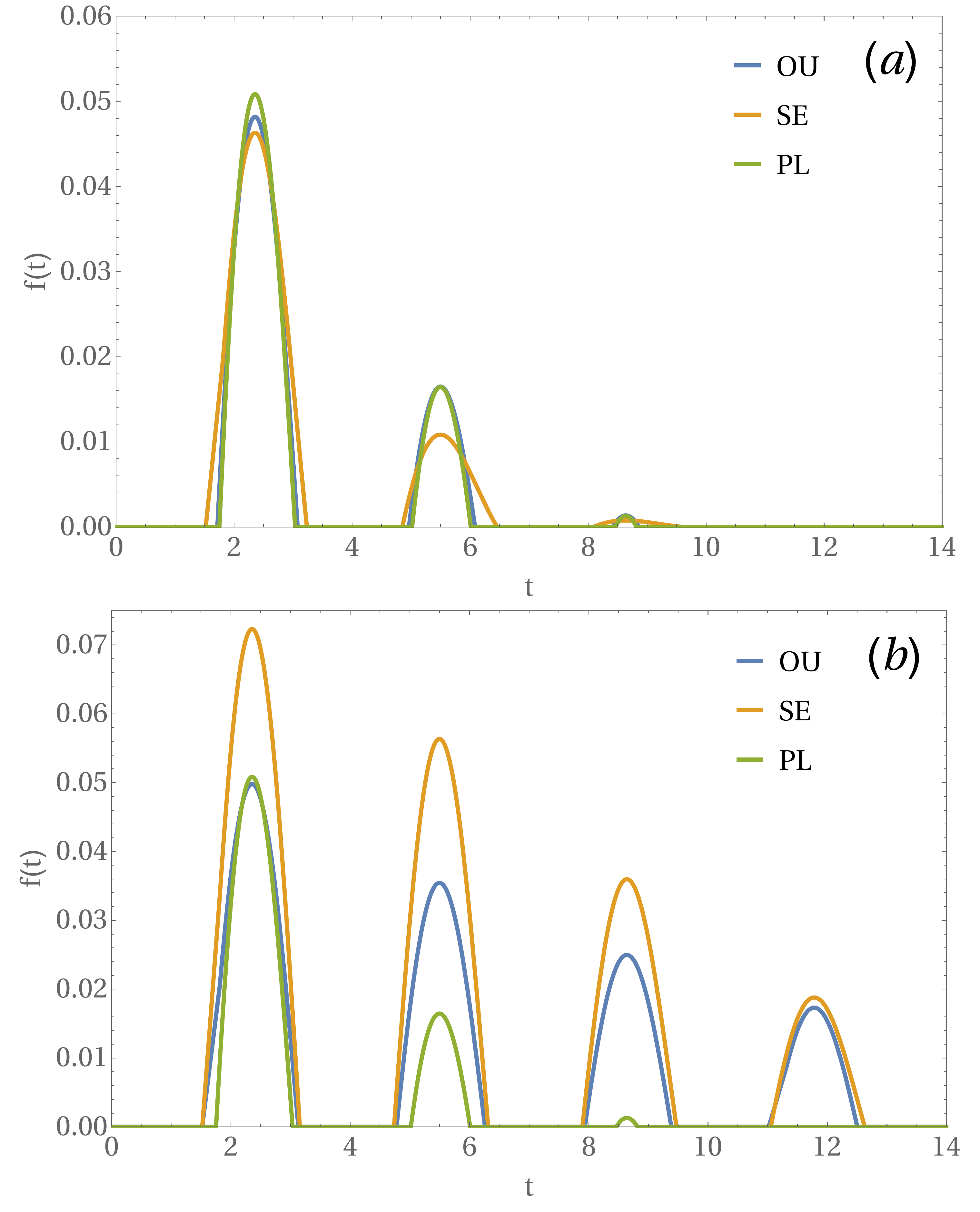}
	\caption{(Color online) ACHL measure $f(t)$ for different noises. (a) The parameters used were: $\Delta\omega=2$, $\tau_c=4.25$, $\Omega=0.91$ (OU); $\Delta\omega=2$, $\tau_c=4.25$, $\Omega=0.47$ (SE); $\Delta\omega=2$, $\tau_c=10$, $\Omega=1.35$ (PL). (b) The parameters used were: $\Delta\omega=2$, $\tau_c=10$, $\Omega=1.35$. All the curves used: $\gamma=1$, $T=0.1\,\omega_0$.}
	\label{fig:ACHLALL_DIFF}
\end{figure}

The decay rates $\gamma_1(t)$ for the three pumps are plotted in Fig.\ref{fig:GAMMA15ALL}(a) and the average energies in Fig.\ref{fig:GAMMA15ALL}(b), for the non-Markovian case ($T=0.1\,\omega_0$). The function $f(t)$, Eq.(\ref{eq:ACHL_f}), for the three studied noises, with the parameters as in previous sections and for the case $T=0.1\,\omega_0$, is shown in Fig.\ref{fig:ACHLALL_DIFF}(a), where the blue, yellow and green lines correspond, respectively, to the exponential, the squared exponential and the power law noises. Note that the ACHL measure, Eq.(\ref{eq:ACHL}), corresponds to the area under the functions. Since it is clearly positive, the non-Markovianity of the evolution is verified. The values of the measure are approximately $\mathcal{N}_{\textup{ACHL}}^{\textup{OU}}=0.0546$, $\mathcal{N}_{\textup{ACHL}}^{\textup{SE}}=0.0585$ and $\mathcal{N}_{\textup{ACHL}}^{\textup{PL}}=0.0532$ for the exponential, squared exponential and power law noises, respectively. In that regime, the three noises are very similar.

For a better comparison among the non-Markovianity generated by the noises, we plotted in Fig.\ref{fig:ACHLALL_DIFF}(b) the ACHL measure using the parameters of the power law case, where we find non-Markovianity for all the three cases. It can be seen that, for the same parameters, the squared exponencial noise shows a bigger value in the non-Markovianity measure than the others. Namely, the values are $\mathcal{N}_{\textup{ACHL}}^{\textup{OU}}=0.1047$, $\mathcal{N}_{\textup{ACHL}}^{\textup{SE}}=0.1651$ and $\mathcal{N}_{\textup{ACHL}}^{\textup{PL}}=0.0532$ for the exponential, squared exponential and power law noises, respectively. In all these cases, however, the other decay rate $\gamma_2(t)$ was always positive, since the noise strength cannot be large enough to overcome the temperature term $\gamma \left [N(\omega)+1\right)]$, as we are in the weak coupling regime. Therefore, although our evolution is non-Markovian in the RHP sense, it is Markovian by the BLP definiton.

\section{Experimental Proposal}
\label{sec:V}

Single trapped ions are great test bench for reservoir engineering. Changes in the trapping potential and laser-ion interactions can be used to create Amplitude and Phase reservoirs \cite{QuantReservEngPoyatos1996,DecoherenceMyatt200,QuantHarmoReservEngKienzler2014}. Another important feature of this system is the ability to do full quantum state tomography \cite{QuantumDynamicsLeibfried2003, QuantumTomographyRoss2004}, which is a  key feature to see the signature of non-Markovian dynamics in the different types of measurements. Therefore, they are a good candidate to test our proposal. 
 
A linear Paul trap combines oscillating and static electric fields to create an effective static 3D harmonic potential. If we considered the radial trapping frequency ($\omega_r$) much higher than the axial one ($\omega$) the ion motion is simplified. In this approximation, the net system comprises of one ion in a 1D harmonic motion. Choosing two internal metastable electronic levels, the ion internal structure can be approximated by a two-level system, which can be represented in the spin $1/2$ basis $ \ket{\uparrow} $ $ \ket{\downarrow}$.

The Hamiltonian that describes the quantum dynamics of the single trapped ion interacting with the light field is:
 
\begin{equation}
 \label{eq:HAMILTON_LaserIonFull}
 H = \frac{\omega_0}{2}\sigma_z + \omega a^{\dagger}a + \frac{\Omega}{2}\left(\sigma_{+} + \sigma_{-}\right)\left[e^{i(kz-\omega_l+\phi)}+ e^{-i(kz-\omega_l+\phi)}\right],
\end{equation}
where $\Omega$ is the coupling strength of the laser, $k$ is the wave number, $\omega_l$ and $\phi$ are the laser frequency and phase, respectively. There are several approximations and considerations that we can take into account \cite{QuantumDynamicsLeibfried2003} that simplify the Hamiltonian and enable us to express it in terms of the harmonic oscillator operators of creation and annihilation:
 
\begin{equation}
 \label{eq:HAMILTON_LaserIonApprox}
 H = \frac{\omega_0}{2}\sigma_z + \omega a^{\dagger}a + \frac{\Omega}{2}\left[ e^{i\eta (a+a^{\dagger})}\sigma_{+}e^{-i(\omega_l+\phi)} + e^{-i\eta (a+a^{\dagger})}\sigma_{-}e^{+i(\omega_l+\phi)} \right],
\end{equation}
where the Lamb-Dicke parameter is defined as 
\begin{equation}
\eta= k \sqrt{\frac{\hbar}{2m_0 \omega}}.
\end{equation}

With laser cooling methods the ion motion can be prepared near to the ground state of the oscillator \cite{QuantumDynamicsLeibfried2003}. This is know as the Lamb-Dicke regime, where $\eta <<1 $. Therefore, we can consider that the average vibrational occupation number is ($ \braket{n} \approx 0$). In this regime the exponential terms can be expanded in powers of $\eta$. Keeping only the terms up to first order $\eta$ the light field ion  interaction is described by three resonant terms: 

\begin{equation}
\label{eq:HAMILTON_LaserIonCarrier}
H_\textup{c} = \frac{1}{2}\,\hbar\, \Omega_n \,n \left(\sigma_{+}\,e^{i\phi}+\sigma_{-}\,e^{i\phi}\right),
\end{equation}
\begin{equation}
\label{eq:HAMILTON_LaserIonRSB}
H_{\textup{rsb}} =  \frac{1}{2}\,\hbar\, \Omega_{\textup{sb}} \left(a\,\sigma_{+}\,e^{i\phi}- a^{\dagger}\sigma_{-}\,e^{i\phi}\right),
\end{equation} 
 \begin{equation}
 \label{eq:HAMILTON_LaserIonBSB}
H_{\textup{bsb}} =  \frac{1}{2}\,\hbar\, \Omega_{\textup{sb}} \left(a^{\dagger}\sigma_{+}\,e^{i\phi}- a\,\sigma_{-}\,e^{i\phi}\right),
\end{equation} 

where $\Omega_c = \Omega\,(1-\eta^2)$, $\Omega_{\textup{sb}} =\eta \,\Omega $ are the coupling strength of the Raman transitions. The first term is the Carrier transition, this coupling changes only the internal degree of freedom, the ion motion is kept intact. The second (third) term couples the internal degree of freedom with the motion. It connects the state $\ket{\downarrow\,, n} \leftrightarrow \ket{\downarrow\,, n-1}$ $\left(\,\ket{\uparrow\,, n} \leftrightarrow \ket{\uparrow\,, n+1}\,\right)$. Therefore, it is know as the Red-Side Band (Blue-Side Band) transition.

We can see that the two-level system coupled to a harmonic bath plus color noise can be fully simulated with a combination of this transitions. Comparing this terms with Eq. (\ref{eq:HALMILTON_ampdamp}) with we  choose the the right phase for the laser, the Blue-Side Band interaction mimics the amplitude damping Hamiltonian, were the environment has only one mode of vibration. The noise is modeled using a stochastic time dependent light field in the Carrier transition. Experimentally this can be achieved with a arbitrary function generator and a acousto-optic modulator to modulate the amplitude and phase of the laser. The Markovian temperature, in this case, will be related to the ratio between the Blue-Side Band and the Carrier laser intensities.

We can go one step further and couple one oscillator system to a oscillator bath, as it was done by Myatt et al.\cite{DecoherenceMyatt200}. They used the superposition of coherent motion states to study decoherence through the coupling to engineered amplitude and phase reservoirs. In the experiment, the amplitude reservoir was created adding a white noise in the trap electric field and the Phase dumping by modulating the trap frequency with that noise. Using that technique they were able to engineer high-temperature and zero-temperature reservoirs. The system was prepared in a superposition of coherent motion states and its coherence as function of the size of the superposition was measured with single-atom interferometry \cite{DecoherenceMyatt200}. Adding a colored noise to the white noise, the non-Markovian dynamics could be seen as a revival of the interference fringes.


\section{Conclusions and Outlook}
\label{sec:VI}

We have investigated the dynamics of a system subjected to a classical colored noise and shown that this pump indeed induces quantum non-Markovianity. We have considered a qubit interacting with a bosonic environment and undergoing a Markovian evolution. After turning on the classical noise, the evolution becomes non-Markovian, as witnessed by the ACHL measure applied on the master equation of the system. We have analyzed three different colored noises, showing that the squared exponential noise is the more efficient one to produce this effect.

The evolution of the system has been decomposed in two parts: system-bath and system-pump evolutions. In the first the qubit exchanges energy with a bosonic bath, and a Markovian master equation has been derived under the Born-Markov setting \cite{Rivas2012,Breuer2007}. For the second, the noise has been modelled by a stochastic Hamiltonian and functional calculus and, together with a weak coupling assumption, has been used in order to derive a master equation \cite{Budini2001}. In this case, we have found that the evolution can become non-Markovian, depending on the noise parameters. Having the master equation, we have been able to evaluate the non-Markovianity of the system evolution by analysis of its decay rates, according to the ACHL measure \cite{Hall2014}.

Three noises have been used in the subsequent analysis: exponential (Ornstein Uhlenbeck), squared exponential and power law noises \cite{Benedetti2014}. The idea has been to subject the system to an environment at $T=0.33\,\omega_0$ and find the minimum values of the parameters of each noise required to reach the Markovianity limit. Physically we have been interested in the mininum noise strength necessary to overcome the decohering effects of the bath and reverse the energy flow. Cooling the environment to $T=0.10\,\omega_0$ with fixed noise parameters, we have been able to measure the non-Markovianity of the system dynamics associated to each noise. The noises have been then compared under the same values of the parameters, with the ACHL measure for the squared exponential noise exhibiting bigger values, which shows that, among the three noises, it can generate more quantum non-Markovianity. The noise pumps also have been found to cause the system's average energy to oscillate and increase from the value they reach without the pump. This increase in value, however, results almost independent of the temperature. Since the energy differences between Markovian and non-Markovian regimes are equivalent to all noise, the energy required to carry from one regime to another is low.

Our results suggest further interesting studies within the context of open quantum systems. For instance, the formalism developed here could be applied to actual quantum systems for investigating to which extent (i) the injection of noise acts as a reliable alternative to reservoir engineering and (ii) the non-Markovian features of the model are relevant to decoherence suppression or preservation of information. Moreover, the model could be exploited for quantum thermal machines \cite{Palao2001,Rezek2009} to verify its possible role in enabling useful thermodynamical features, such as increased efficiencies of cycles.

\section*{Acknowledgements}
\label{sec:VII}

The authors would like to thank G. Adesso, F. F. Fanchini, F. Brito, and M. H. Y. Moussa for fruitful conversations. The authors would like to acknowledge the financial support from Brazilian funding agencies CNPq (Grants No. 304955/2013-2, 443828/2014-8), CAPES Pesquisador Visitante Especial (Grant No. 108/2012), FAPESP (Grant No. 2014/09566-0) and the Brazilian National Institute of Science and Technology of Quantum Information (INCT/IQ).


\begin{thebibliography}{88}%
\makeatletter
\providecommand \@ifxundefined [1]{%
 \@ifx{#1\undefined}
}%
\providecommand \@ifnum [1]{%
 \ifnum #1\expandafter \@firstoftwo
 \else \expandafter \@secondoftwo
 \fi
}%
\providecommand \@ifx [1]{%
 \ifx #1\expandafter \@firstoftwo
 \else \expandafter \@secondoftwo
 \fi
}%
\providecommand \natexlab [1]{#1}%
\providecommand \enquote  [1]{``#1''}%
\providecommand \bibnamefont  [1]{#1}%
\providecommand \bibfnamefont [1]{#1}%
\providecommand \citenamefont [1]{#1}%
\providecommand \href@noop [0]{\@secondoftwo}%
\providecommand \href [0]{\begingroup \@sanitize@url \@href}%
\providecommand \@href[1]{\@@startlink{#1}\@@href}%
\providecommand \@@href[1]{\endgroup#1\@@endlink}%
\providecommand \@sanitize@url [0]{\catcode `\\12\catcode `\$12\catcode
  `\&12\catcode `\#12\catcode `\^12\catcode `\_12\catcode `\%12\relax}%
\providecommand \@@startlink[1]{}%
\providecommand \@@endlink[0]{}%
\providecommand \url  [0]{\begingroup\@sanitize@url \@url }%
\providecommand \@url [1]{\endgroup\@href {#1}{\urlprefix }}%
\providecommand \urlprefix  [0]{URL }%
\providecommand \Eprint [0]{\href }%
\providecommand \doibase [0]{http://dx.doi.org/}%
\providecommand \selectlanguage [0]{\@gobble}%
\providecommand \bibinfo  [0]{\@secondoftwo}%
\providecommand \bibfield  [0]{\@secondoftwo}%
\providecommand \translation [1]{[#1]}%
\providecommand \BibitemOpen [0]{}%
\providecommand \bibitemStop [0]{}%
\providecommand \bibitemNoStop [0]{.\EOS\space}%
\providecommand \EOS [0]{\spacefactor3000\relax}%
\providecommand \BibitemShut  [1]{\csname bibitem#1\endcsname}%
\let\auto@bib@innerbib\@empty
\bibitem [{\citenamefont {Breuer}\ and\ \citenamefont
  {Petruccione}(2007)}]{Breuer2007}%
  \BibitemOpen
  \bibfield  {author} {\bibinfo {author} {\bibfnamefont {H.-P.}\ \bibnamefont
  {Breuer}}\ and\ \bibinfo {author} {\bibfnamefont {F.}~\bibnamefont
  {Petruccione}},\ }\href
  {http://dx.doi.org/10.1093/acprof:oso/9780199213900.001.0001} {\emph
  {\bibinfo {title} {The Theory of Open Quantum Systems}}}\ (\bibinfo
  {publisher} {Oxford University Press ({OUP})},\ \bibinfo {year}
  {2007})\BibitemShut {NoStop}%
\bibitem [{\citenamefont {Rivas}\ and\ \citenamefont
  {Huelga}(2012)}]{Rivas2012}%
  \BibitemOpen
  \bibfield  {author} {\bibinfo {author} {\bibfnamefont {A.}~\bibnamefont
  {Rivas}}\ and\ \bibinfo {author} {\bibfnamefont {S.~F.}\ \bibnamefont
  {Huelga}},\ }\href {http://dx.doi.org/10.1007/978-3-642-23354-8} {\emph
  {\bibinfo {title} {Open Quantum Systems}}}\ (\bibinfo  {publisher} {Springer
  Berlin Heidelberg},\ \bibinfo {year} {2012})\BibitemShut {NoStop}%
\bibitem [{\citenamefont {Carmichael}(1993)}]{Carmichael199305}%
  \BibitemOpen
  \bibfield  {author} {\bibinfo {author} {\bibfnamefont {H.}~\bibnamefont
  {Carmichael}},\ }\href {http://amazon.com/o/ASIN/3540566341/} {\emph
  {\bibinfo {title} {An Open Systems Approach to Quantum Optics (Lecture Notes
  in Physics)}}}\ (\bibinfo  {publisher} {Springer},\ \bibinfo {year}
  {1993})\BibitemShut {NoStop}%
\bibitem [{ALI(2007)}]{ALICKI}%
  \BibitemOpen
  \href {http://dx.doi.org/10.1007/3-540-70861-8} {\emph {\bibinfo {title}
  {Quantum Dynamical Semigroups and Applications}}}\ (\bibinfo  {publisher}
  {Springer Berlin Heidelberg},\ \bibinfo {year} {2007})\BibitemShut {NoStop}%
\bibitem [{\citenamefont {Davis}(1976)}]{Davis197611}%
  \BibitemOpen
  \bibfield  {author} {\bibinfo {author} {\bibfnamefont {E.}~\bibnamefont
  {Davis}},\ }\href {http://amazon.com/o/ASIN/0122061500/} {\emph {\bibinfo
  {title} {Quantum Theory of Open Systems}}}\ (\bibinfo  {publisher} {Academic
  Press Inc},\ \bibinfo {year} {1976})\BibitemShut {NoStop}%
\bibitem [{\citenamefont {Weiss}(2012)}]{Weiss201203}%
  \BibitemOpen
  \bibfield  {author} {\bibinfo {author} {\bibfnamefont {U.}~\bibnamefont
  {Weiss}},\ }\href {http://amazon.com/o/ASIN/9814374911/} {\emph {\bibinfo
  {title} {Quantum Dissipative Systems (4th Edition)}}},\ \bibinfo {edition}
  {4th}\ ed.\ (\bibinfo  {publisher} {World Scientific Publishing Company},\
  \bibinfo {year} {2012})\BibitemShut {NoStop}%
\bibitem [{\citenamefont {Joos}\ \emph {et~al.}(2003)\citenamefont {Joos},
  \citenamefont {Zeh}, \citenamefont {Kiefer}, \citenamefont {Giulini},
  \citenamefont {Kupsch},\ and\ \citenamefont {Stamatescu}}]{Joos2003}%
  \BibitemOpen
  \bibfield  {author} {\bibinfo {author} {\bibfnamefont {E.}~\bibnamefont
  {Joos}}, \bibinfo {author} {\bibfnamefont {H.~D.}\ \bibnamefont {Zeh}},
  \bibinfo {author} {\bibfnamefont {C.}~\bibnamefont {Kiefer}}, \bibinfo
  {author} {\bibfnamefont {D.}~\bibnamefont {Giulini}}, \bibinfo {author}
  {\bibfnamefont {J.}~\bibnamefont {Kupsch}}, \ and\ \bibinfo {author}
  {\bibfnamefont {I.-O.}\ \bibnamefont {Stamatescu}},\ }\href
  {http://dx.doi.org/10.1007/978-3-662-05328-7} {\emph {\bibinfo {title}
  {Decoherence and the Appearance of a Classical World in Quantum Theory}}}\
  (\bibinfo  {publisher} {Springer Science Business Media},\ \bibinfo {year}
  {2003})\BibitemShut {NoStop}%
\bibitem [{\citenamefont {Zurek}(2003)}]{Zurek2003}%
  \BibitemOpen
  \bibfield  {author} {\bibinfo {author} {\bibfnamefont {W.~H.}\ \bibnamefont
  {Zurek}},\ }\href {http://dx.doi.org/10.1103/RevModPhys.75.715} {\bibfield
  {journal} {\bibinfo  {journal} {Rev. Mod. Phys.}\ }\textbf {\bibinfo {volume}
  {75}},\ \bibinfo {pages} {715} (\bibinfo {year} {2003})}\BibitemShut
  {NoStop}%
\bibitem [{\citenamefont {Rivas}\ \emph {et~al.}(2014)\citenamefont {Rivas},
  \citenamefont {Huelga},\ and\ \citenamefont {Plenio}}]{Rivas2014}%
  \BibitemOpen
  \bibfield  {author} {\bibinfo {author} {\bibfnamefont {{\'{A}}.}~\bibnamefont
  {Rivas}}, \bibinfo {author} {\bibfnamefont {S.~F.}\ \bibnamefont {Huelga}}, \
  and\ \bibinfo {author} {\bibfnamefont {M.~B.}\ \bibnamefont {Plenio}},\
  }\href {http://dx.doi.org/10.1088/0034-4885/77/9/094001} {\bibfield
  {journal} {\bibinfo  {journal} {Rep. Prog. Phys.}\ }\textbf {\bibinfo
  {volume} {77}},\ \bibinfo {pages} {094001} (\bibinfo {year}
  {2014})}\BibitemShut {NoStop}%
\bibitem [{\citenamefont {Breuer}\ \emph {et~al.}(2016)\citenamefont {Breuer},
  \citenamefont {Laine}, \citenamefont {Piilo},\ and\ \citenamefont
  {Vacchini}}]{Breuer2016}%
  \BibitemOpen
  \bibfield  {author} {\bibinfo {author} {\bibfnamefont {H.-P.}\ \bibnamefont
  {Breuer}}, \bibinfo {author} {\bibfnamefont {E.-M.}\ \bibnamefont {Laine}},
  \bibinfo {author} {\bibfnamefont {J.}~\bibnamefont {Piilo}}, \ and\ \bibinfo
  {author} {\bibfnamefont {B.}~\bibnamefont {Vacchini}},\ }\href
  {http://dx.doi.org/10.1103/RevModPhys.88.021002} {\bibfield  {journal}
  {\bibinfo  {journal} {Rev. Mod. Phys.}\ }\textbf {\bibinfo {volume} {88}},\
  \bibinfo {pages} {021002} (\bibinfo {year} {2016})}\BibitemShut {NoStop}%
\bibitem [{\citenamefont {Wolf}\ \emph {et~al.}(2008)\citenamefont {Wolf},
  \citenamefont {Eisert}, \citenamefont {Cubitt},\ and\ \citenamefont
  {Cirac}}]{Wolf2008}%
  \BibitemOpen
  \bibfield  {author} {\bibinfo {author} {\bibfnamefont {M.~M.}\ \bibnamefont
  {Wolf}}, \bibinfo {author} {\bibfnamefont {J.}~\bibnamefont {Eisert}},
  \bibinfo {author} {\bibfnamefont {T.~S.}\ \bibnamefont {Cubitt}}, \ and\
  \bibinfo {author} {\bibfnamefont {J.~I.}\ \bibnamefont {Cirac}},\ }\href
  {http://dx.doi.org/10.1103/PhysRevLett.101.150402} {\bibfield  {journal}
  {\bibinfo  {journal} {Phys. Rev. Lett.}\ }\textbf {\bibinfo {volume} {101}},\
  \bibinfo {pages} {150402} (\bibinfo {year} {2008})}\BibitemShut {NoStop}%
\bibitem [{\citenamefont {Piilo}\ \emph {et~al.}(2008)\citenamefont {Piilo},
  \citenamefont {Maniscalco}, \citenamefont {H\"{a}rk\"{o}nen},\ and\
  \citenamefont {Suominen}}]{Piilo2008}%
  \BibitemOpen
  \bibfield  {author} {\bibinfo {author} {\bibfnamefont {J.}~\bibnamefont
  {Piilo}}, \bibinfo {author} {\bibfnamefont {S.}~\bibnamefont {Maniscalco}},
  \bibinfo {author} {\bibfnamefont {K.}~\bibnamefont {H\"{a}rk\"{o}nen}}, \
  and\ \bibinfo {author} {\bibfnamefont {K.-A.}\ \bibnamefont {Suominen}},\
  }\href {http://dx.doi.org/10.1103/PhysRevLett.100.180402} {\bibfield
  {journal} {\bibinfo  {journal} {Phys. Rev. Lett.}\ }\textbf {\bibinfo
  {volume} {100}},\ \bibinfo {pages} {180402} (\bibinfo {year}
  {2008})}\BibitemShut {NoStop}%
\bibitem [{\citenamefont {{de Vega}}\ and\ \citenamefont
  {{Alonso}}(2015)}]{Vega2015}%
  \BibitemOpen
  \bibfield  {author} {\bibinfo {author} {\bibfnamefont {I.}~\bibnamefont {{de
  Vega}}}\ and\ \bibinfo {author} {\bibfnamefont {D.}~\bibnamefont
  {{Alonso}}},\ }\href@noop {} {\  (\bibinfo {year} {2015})},\ \Eprint
  {http://arxiv.org/abs/1511.06994} {arXiv:1511.06994 [quant-ph]} \BibitemShut
  {NoStop}%
\bibitem [{\citenamefont {{Pollock}}\ \emph {et~al.}(2015)\citenamefont
  {{Pollock}}, \citenamefont {{Rodr{\'{\i}}guez-Rosario}}, \citenamefont
  {{Frauenheim}}, \citenamefont {{Paternostro}},\ and\ \citenamefont
  {{Modi}}}]{Modi2015}%
  \BibitemOpen
  \bibfield  {author} {\bibinfo {author} {\bibfnamefont {F.~A.}\ \bibnamefont
  {{Pollock}}}, \bibinfo {author} {\bibfnamefont {C.}~\bibnamefont
  {{Rodr{\'{\i}}guez-Rosario}}}, \bibinfo {author} {\bibfnamefont
  {T.}~\bibnamefont {{Frauenheim}}}, \bibinfo {author} {\bibfnamefont
  {M.}~\bibnamefont {{Paternostro}}}, \ and\ \bibinfo {author} {\bibfnamefont
  {K.}~\bibnamefont {{Modi}}},\ }\href@noop {} {\  (\bibinfo {year} {2015})},\
  \Eprint {http://arxiv.org/abs/1512.00589} {arXiv:1512.00589 [quant-ph]}
  \BibitemShut {NoStop}%
\bibitem [{\citenamefont {Brito}\ and\ \citenamefont
  {Werlang}(2015)}]{Brito2015}%
  \BibitemOpen
  \bibfield  {author} {\bibinfo {author} {\bibfnamefont {F.}~\bibnamefont
  {Brito}}\ and\ \bibinfo {author} {\bibfnamefont {T.}~\bibnamefont
  {Werlang}},\ }\href {http://dx.doi.org/10.1088/1367-2630/17/7/072001}
  {\bibfield  {journal} {\bibinfo  {journal} {New Journal of Physics}\ }\textbf
  {\bibinfo {volume} {17}},\ \bibinfo {pages} {072001} (\bibinfo {year}
  {2015})}\BibitemShut {NoStop}%
\bibitem [{\citenamefont {Rivas}\ \emph {et~al.}(2010)\citenamefont {Rivas},
  \citenamefont {Huelga},\ and\ \citenamefont {Plenio}}]{Rivas2010}%
  \BibitemOpen
  \bibfield  {author} {\bibinfo {author} {\bibfnamefont {{\'{A}}.}~\bibnamefont
  {Rivas}}, \bibinfo {author} {\bibfnamefont {S.~F.}\ \bibnamefont {Huelga}}, \
  and\ \bibinfo {author} {\bibfnamefont {M.~B.}\ \bibnamefont {Plenio}},\
  }\href {http://dx.doi.org/10.1103/PhysRevLett.105.050403} {\bibfield
  {journal} {\bibinfo  {journal} {Phys. Rev. Lett.}\ }\textbf {\bibinfo
  {volume} {105}},\ \bibinfo {pages} {050403} (\bibinfo {year}
  {2010})}\BibitemShut {NoStop}%
\bibitem [{\citenamefont {Bylicka}\ \emph {et~al.}(2014)\citenamefont
  {Bylicka}, \citenamefont {Chru{\'{s}}ci{\'{n}}ski},\ and\ \citenamefont
  {Maniscalco}}]{Bylicka2014}%
  \BibitemOpen
  \bibfield  {author} {\bibinfo {author} {\bibfnamefont {B.}~\bibnamefont
  {Bylicka}}, \bibinfo {author} {\bibfnamefont {D.}~\bibnamefont
  {Chru{\'{s}}ci{\'{n}}ski}}, \ and\ \bibinfo {author} {\bibfnamefont
  {S.}~\bibnamefont {Maniscalco}},\ }\href
  {http://dx.doi.org/10.1038/srep05720} {\bibfield  {journal} {\bibinfo
  {journal} {Sci. Rep.}\ }\textbf {\bibinfo {volume} {4}},\ \bibinfo {pages}
  {5720} (\bibinfo {year} {2014})}\BibitemShut {NoStop}%
\bibitem [{\citenamefont {Addis}\ \emph {et~al.}(2014)\citenamefont {Addis},
  \citenamefont {Bylicka}, \citenamefont {Chru{\'{s}}ci{\'{n}}ski},\ and\
  \citenamefont {Maniscalco}}]{Addis2014}%
  \BibitemOpen
  \bibfield  {author} {\bibinfo {author} {\bibfnamefont {C.}~\bibnamefont
  {Addis}}, \bibinfo {author} {\bibfnamefont {B.}~\bibnamefont {Bylicka}},
  \bibinfo {author} {\bibfnamefont {D.}~\bibnamefont
  {Chru{\'{s}}ci{\'{n}}ski}}, \ and\ \bibinfo {author} {\bibfnamefont
  {S.}~\bibnamefont {Maniscalco}},\ }\href
  {http://dx.doi.org/10.1103/PhysRevA.90.052103} {\bibfield  {journal}
  {\bibinfo  {journal} {Phys. Rev. A}\ }\textbf {\bibinfo {volume} {90}},\
  \bibinfo {pages} {052103} (\bibinfo {year} {2014})}\BibitemShut {NoStop}%
\bibitem [{\citenamefont {Leggio}\ \emph {et~al.}(2015)\citenamefont {Leggio},
  \citenamefont {{Lo Franco}}, \citenamefont {Soares-Pinto}, \citenamefont
  {Horodecki},\ and\ \citenamefont {Compagno}}]{Leggio2015}%
  \BibitemOpen
  \bibfield  {author} {\bibinfo {author} {\bibfnamefont {B.}~\bibnamefont
  {Leggio}}, \bibinfo {author} {\bibfnamefont {R.}~\bibnamefont {{Lo Franco}}},
  \bibinfo {author} {\bibfnamefont {D.~O.}\ \bibnamefont {Soares-Pinto}},
  \bibinfo {author} {\bibfnamefont {P.}~\bibnamefont {Horodecki}}, \ and\
  \bibinfo {author} {\bibfnamefont {G.}~\bibnamefont {Compagno}},\ }\href
  {http://dx.doi.org/10.1103/PhysRevA.92.032311} {\bibfield  {journal}
  {\bibinfo  {journal} {Phys. Rev. A}\ }\textbf {\bibinfo {volume} {92}},\
  \bibinfo {pages} {032311} (\bibinfo {year} {2015})}\BibitemShut {NoStop}%
\bibitem [{\citenamefont {Lloyd}\ and\ \citenamefont
  {Viola}(2001)}]{Lloyd2001}%
  \BibitemOpen
  \bibfield  {author} {\bibinfo {author} {\bibfnamefont {S.}~\bibnamefont
  {Lloyd}}\ and\ \bibinfo {author} {\bibfnamefont {L.}~\bibnamefont {Viola}},\
  }\href {http://dx.doi.org/10.1103/PhysRevA.65.010101} {\bibfield  {journal}
  {\bibinfo  {journal} {Phys. Rev. A}\ }\textbf {\bibinfo {volume} {65}},\
  \bibinfo {pages} {010101} (\bibinfo {year} {2001})}\BibitemShut {NoStop}%
\bibitem [{\citenamefont {Chin}\ \emph {et~al.}(2012)\citenamefont {Chin},
  \citenamefont {Huelga},\ and\ \citenamefont {Plenio}}]{Chin2012}%
  \BibitemOpen
  \bibfield  {author} {\bibinfo {author} {\bibfnamefont {A.~W.}\ \bibnamefont
  {Chin}}, \bibinfo {author} {\bibfnamefont {S.~F.}\ \bibnamefont {Huelga}}, \
  and\ \bibinfo {author} {\bibfnamefont {M.~B.}\ \bibnamefont {Plenio}},\
  }\href {http://dx.doi.org/10.1103/PhysRevLett.109.233601} {\bibfield
  {journal} {\bibinfo  {journal} {Phys. Rev. Lett.}\ }\textbf {\bibinfo
  {volume} {109}},\ \bibinfo {pages} {233601} (\bibinfo {year}
  {2012})}\BibitemShut {NoStop}%
\bibitem [{\citenamefont {Verstraete}\ \emph {et~al.}(2009)\citenamefont
  {Verstraete}, \citenamefont {Wolf},\ and\ \citenamefont
  {Cirac}}]{Verstraete2009}%
  \BibitemOpen
  \bibfield  {author} {\bibinfo {author} {\bibfnamefont {F.}~\bibnamefont
  {Verstraete}}, \bibinfo {author} {\bibfnamefont {M.~M.}\ \bibnamefont
  {Wolf}}, \ and\ \bibinfo {author} {\bibfnamefont {J.~I.}\ \bibnamefont
  {Cirac}},\ }\href {http://dx.doi.org/10.1038/nphys1342} {\bibfield  {journal}
  {\bibinfo  {journal} {Nat. Phys.}\ }\textbf {\bibinfo {volume} {5}},\
  \bibinfo {pages} {633} (\bibinfo {year} {2009})}\BibitemShut {NoStop}%
\bibitem [{\citenamefont {Man}\ \emph {et~al.}(2015{\natexlab{a}})\citenamefont
  {Man}, \citenamefont {Xia},\ and\ \citenamefont {{Lo
  Franco}}}]{man2015cavity}%
  \BibitemOpen
  \bibfield  {author} {\bibinfo {author} {\bibfnamefont {Z.-X.}\ \bibnamefont
  {Man}}, \bibinfo {author} {\bibfnamefont {Y.-J.}\ \bibnamefont {Xia}}, \ and\
  \bibinfo {author} {\bibfnamefont {R.}~\bibnamefont {{Lo Franco}}},\
  }\href@noop {} {\bibfield  {journal} {\bibinfo  {journal} {Sci. Rep.}\
  }\textbf {\bibinfo {volume} {5}},\ \bibinfo {pages} {13843} (\bibinfo {year}
  {2015}{\natexlab{a}})}\BibitemShut {NoStop}%
\bibitem [{\citenamefont {Man}\ \emph {et~al.}(2015{\natexlab{b}})\citenamefont
  {Man}, \citenamefont {Xia},\ and\ \citenamefont {{Lo
  Franco}}}]{man2015harnessing}%
  \BibitemOpen
  \bibfield  {author} {\bibinfo {author} {\bibfnamefont {Z.-X.}\ \bibnamefont
  {Man}}, \bibinfo {author} {\bibfnamefont {Y.-J.}\ \bibnamefont {Xia}}, \ and\
  \bibinfo {author} {\bibfnamefont {R.}~\bibnamefont {{Lo Franco}}},\
  }\href@noop {} {\bibfield  {journal} {\bibinfo  {journal} {Phys. Rev. A}\
  }\textbf {\bibinfo {volume} {92}},\ \bibinfo {pages} {012315} (\bibinfo
  {year} {2015}{\natexlab{b}})}\BibitemShut {NoStop}%
\bibitem [{\citenamefont {Romero}\ and\ \citenamefont {{Lo
  Franco}}(2012)}]{romero2012simple}%
  \BibitemOpen
  \bibfield  {author} {\bibinfo {author} {\bibfnamefont {K.~F.}\ \bibnamefont
  {Romero}}\ and\ \bibinfo {author} {\bibfnamefont {R.}~\bibnamefont {{Lo
  Franco}}},\ }\href@noop {} {\bibfield  {journal} {\bibinfo  {journal} {Phys.
  Scr.}\ }\textbf {\bibinfo {volume} {86}},\ \bibinfo {pages} {065004}
  (\bibinfo {year} {2012})}\BibitemShut {NoStop}%
\bibitem [{\citenamefont {{Lo Franco}}(2015)}]{lo2015switching}%
  \BibitemOpen
  \bibfield  {author} {\bibinfo {author} {\bibfnamefont {R.}~\bibnamefont {{Lo
  Franco}}},\ }\href@noop {} {\bibfield  {journal} {\bibinfo  {journal} {New J.
  Phys.}\ }\textbf {\bibinfo {volume} {17}},\ \bibinfo {pages} {081004}
  (\bibinfo {year} {2015})}\BibitemShut {NoStop}%
\bibitem [{\citenamefont {{Lo Franco}}(2016)}]{lo2016nonlocality}%
  \BibitemOpen
  \bibfield  {author} {\bibinfo {author} {\bibfnamefont {R.}~\bibnamefont {{Lo
  Franco}}},\ }\href@noop {} {\bibfield  {journal} {\bibinfo  {journal}
  {Quantum Inform. Process.}\ }\textbf {\bibinfo {volume} {15}},\ \bibinfo
  {pages} {2393} (\bibinfo {year} {2016})}\BibitemShut {NoStop}%
\bibitem [{\citenamefont {Gonz{\'{a}}lez-Guti{\'{e}}rrez}\ \emph
  {et~al.}(2016)\citenamefont {Gonz{\'{a}}lez-Guti{\'{e}}rrez}, \citenamefont
  {Rom{\'{a}}n-Ancheyta}, \citenamefont {Espitia},\ and\ \citenamefont {{Lo
  Franco}}}]{GonzlezGutirrez2016}%
  \BibitemOpen
  \bibfield  {author} {\bibinfo {author} {\bibfnamefont {C.~A.}\ \bibnamefont
  {Gonz{\'{a}}lez-Guti{\'{e}}rrez}}, \bibinfo {author} {\bibfnamefont
  {R.}~\bibnamefont {Rom{\'{a}}n-Ancheyta}}, \bibinfo {author} {\bibfnamefont
  {D.}~\bibnamefont {Espitia}}, \ and\ \bibinfo {author} {\bibfnamefont
  {R.}~\bibnamefont {{Lo Franco}}},\ }\href
  {http://dx.doi.org/10.1142/S0219749916500313} {\bibfield  {journal} {\bibinfo
   {journal} {Int. J. Quantum Inform.}\ }\textbf {\bibinfo {volume} {14}},\
  \bibinfo {pages} {1650031} (\bibinfo {year} {2016})}\BibitemShut {NoStop}%
\bibitem [{\citenamefont {{van Kampen}}(2007)}]{Kampen200705}%
  \BibitemOpen
  \bibfield  {author} {\bibinfo {author} {\bibfnamefont {N.}~\bibnamefont {{van
  Kampen}}},\ }\href {http://amazon.com/o/ASIN/0444529659/} {\emph {\bibinfo
  {title} {Stochastic Processes in Physics and Chemistry, Third Edition
  (North-Holland Personal Library)}}},\ \bibinfo {edition} {3rd}\ ed.\
  (\bibinfo  {publisher} {North Holland},\ \bibinfo {year} {2007})\BibitemShut
  {NoStop}%
\bibitem [{\citenamefont {Accardi}(1985)}]{Accardi1985}%
  \BibitemOpen
  \bibfield  {author} {\bibinfo {author} {\bibfnamefont {L.}~\bibnamefont
  {Accardi}},\ }in\ \href {http://dx.doi.org/10.1007/978-1-4899-6653-7_16}
  {\emph {\bibinfo {booktitle} {Statistical Physics and Dynamical Systems}}}\
  (\bibinfo  {publisher} {Springer Science Business Media},\ \bibinfo {year}
  {1985})\ pp.\ \bibinfo {pages} {285--302}\BibitemShut {NoStop}%
\bibitem [{\citenamefont {Haikka}\ \emph {et~al.}(2011)\citenamefont {Haikka},
  \citenamefont {Cresser},\ and\ \citenamefont {Maniscalco}}]{Haikka2011}%
  \BibitemOpen
  \bibfield  {author} {\bibinfo {author} {\bibfnamefont {P.}~\bibnamefont
  {Haikka}}, \bibinfo {author} {\bibfnamefont {J.~D.}\ \bibnamefont {Cresser}},
  \ and\ \bibinfo {author} {\bibfnamefont {S.}~\bibnamefont {Maniscalco}},\
  }\href {http://dx.doi.org/10.1103/PhysRevA.83.012112} {\bibfield  {journal}
  {\bibinfo  {journal} {Phys. Rev. A}\ }\textbf {\bibinfo {volume} {83}},\
  \bibinfo {pages} {012112} (\bibinfo {year} {2011})}\BibitemShut {NoStop}%
\bibitem [{\citenamefont {Neto}\ \emph {et~al.}(2016)\citenamefont {Neto},
  \citenamefont {Karpat},\ and\ \citenamefont {Fanchini}}]{Neto2016}%
  \BibitemOpen
  \bibfield  {author} {\bibinfo {author} {\bibfnamefont {A.~C.}\ \bibnamefont
  {Neto}}, \bibinfo {author} {\bibfnamefont {G.}~\bibnamefont {Karpat}}, \ and\
  \bibinfo {author} {\bibfnamefont {F.~F.}\ \bibnamefont {Fanchini}},\ }\href
  {http://dx.doi.org/10.1103/PhysRevA.94.032105} {\bibfield  {journal}
  {\bibinfo  {journal} {Phys. Rev. A}\ }\textbf {\bibinfo {volume} {94}},\
  \bibinfo {pages} {032105} (\bibinfo {year} {2016})}\BibitemShut {NoStop}%
\bibitem [{\citenamefont {Breuer}\ \emph {et~al.}(2009)\citenamefont {Breuer},
  \citenamefont {Laine},\ and\ \citenamefont {Piilo}}]{Breuer2009}%
  \BibitemOpen
  \bibfield  {author} {\bibinfo {author} {\bibfnamefont {H.-P.}\ \bibnamefont
  {Breuer}}, \bibinfo {author} {\bibfnamefont {E.-M.}\ \bibnamefont {Laine}}, \
  and\ \bibinfo {author} {\bibfnamefont {J.}~\bibnamefont {Piilo}},\ }\href
  {http://dx.doi.org/10.1103/PhysRevLett.103.210401} {\bibfield  {journal}
  {\bibinfo  {journal} {Phys. Rev. Lett.}\ }\textbf {\bibinfo {volume} {103}},\
  \bibinfo {pages} {210401} (\bibinfo {year} {2009})}\BibitemShut {NoStop}%
\bibitem [{\citenamefont {Viola}\ \emph {et~al.}(1999)\citenamefont {Viola},
  \citenamefont {Knill},\ and\ \citenamefont {Lloyd}}]{Viola1999}%
  \BibitemOpen
  \bibfield  {author} {\bibinfo {author} {\bibfnamefont {L.}~\bibnamefont
  {Viola}}, \bibinfo {author} {\bibfnamefont {E.}~\bibnamefont {Knill}}, \ and\
  \bibinfo {author} {\bibfnamefont {S.}~\bibnamefont {Lloyd}},\ }\href
  {http://dx.doi.org/10.1103/PhysRevLett.82.2417} {\bibfield  {journal}
  {\bibinfo  {journal} {Phys. Rev. Lett.}\ }\textbf {\bibinfo {volume} {82}},\
  \bibinfo {pages} {2417} (\bibinfo {year} {1999})}\BibitemShut {NoStop}%
\bibitem [{\citenamefont {Guarnieri}\ \emph {et~al.}(2016)\citenamefont
  {Guarnieri}, \citenamefont {Uchiyama},\ and\ \citenamefont
  {Vacchini}}]{Guarnieri2016}%
  \BibitemOpen
  \bibfield  {author} {\bibinfo {author} {\bibfnamefont {G.}~\bibnamefont
  {Guarnieri}}, \bibinfo {author} {\bibfnamefont {C.}~\bibnamefont {Uchiyama}},
  \ and\ \bibinfo {author} {\bibfnamefont {B.}~\bibnamefont {Vacchini}},\
  }\href {http://dx.doi.org/10.1103/PhysRevA.93.012118} {\bibfield  {journal}
  {\bibinfo  {journal} {Phys. Rev. A}\ }\textbf {\bibinfo {volume} {93}},\
  \bibinfo {pages} {012118} (\bibinfo {year} {2016})}\BibitemShut {NoStop}%
\bibitem [{\citenamefont {Deffner}\ and\ \citenamefont
  {Lutz}(2013)}]{Deffner2013}%
  \BibitemOpen
  \bibfield  {author} {\bibinfo {author} {\bibfnamefont {S.}~\bibnamefont
  {Deffner}}\ and\ \bibinfo {author} {\bibfnamefont {E.}~\bibnamefont {Lutz}},\
  }\href {http://dx.doi.org/10.1103/PhysRevLett.111.010402} {\bibfield
  {journal} {\bibinfo  {journal} {Phys. Rev. Lett.}\ }\textbf {\bibinfo
  {volume} {111}},\ \bibinfo {pages} {010402} (\bibinfo {year}
  {2013})}\BibitemShut {NoStop}%
\bibitem [{\citenamefont {{Pezzutto}}\ \emph {et~al.}(2016)\citenamefont
  {{Pezzutto}}, \citenamefont {{Paternostro}},\ and\ \citenamefont
  {{Omar}}}]{Pezzutto2016}%
  \BibitemOpen
  \bibfield  {author} {\bibinfo {author} {\bibfnamefont {M.}~\bibnamefont
  {{Pezzutto}}}, \bibinfo {author} {\bibfnamefont {M.}~\bibnamefont
  {{Paternostro}}}, \ and\ \bibinfo {author} {\bibfnamefont {Y.}~\bibnamefont
  {{Omar}}},\ }\href@noop {} {\  (\bibinfo {year} {2016})},\ \Eprint
  {http://arxiv.org/abs/1608.03497} {arXiv:1608.03497 [quant-ph]} \BibitemShut
  {NoStop}%
\bibitem [{\citenamefont {Huelga}\ \emph {et~al.}(2012)\citenamefont {Huelga},
  \citenamefont {Rivas},\ and\ \citenamefont {Plenio}}]{Huelga2012}%
  \BibitemOpen
  \bibfield  {author} {\bibinfo {author} {\bibfnamefont {S.~F.}\ \bibnamefont
  {Huelga}}, \bibinfo {author} {\bibfnamefont {{\'{A}}.}~\bibnamefont {Rivas}},
  \ and\ \bibinfo {author} {\bibfnamefont {M.~B.}\ \bibnamefont {Plenio}},\
  }\href {http://dx.doi.org/10.1103/PhysRevLett.108.160402} {\bibfield
  {journal} {\bibinfo  {journal} {Phys. Rev. Lett.}\ }\textbf {\bibinfo
  {volume} {108}},\ \bibinfo {pages} {160402} (\bibinfo {year}
  {2012})}\BibitemShut {NoStop}%
\bibitem [{\citenamefont {{Lo Franco}}\ \emph {et~al.}(2013)\citenamefont {{Lo
  Franco}}, \citenamefont {Bellomo}, \citenamefont {Maniscalco},\ and\
  \citenamefont {Compagno}}]{lo2013dynamics}%
  \BibitemOpen
  \bibfield  {author} {\bibinfo {author} {\bibfnamefont {R.}~\bibnamefont {{Lo
  Franco}}}, \bibinfo {author} {\bibfnamefont {B.}~\bibnamefont {Bellomo}},
  \bibinfo {author} {\bibfnamefont {S.}~\bibnamefont {Maniscalco}}, \ and\
  \bibinfo {author} {\bibfnamefont {G.}~\bibnamefont {Compagno}},\ }\href@noop
  {} {\bibfield  {journal} {\bibinfo  {journal} {Int. J. Mod. Phys. B}\
  }\textbf {\bibinfo {volume} {27}},\ \bibinfo {pages} {1345053} (\bibinfo
  {year} {2013})}\BibitemShut {NoStop}%
\bibitem [{\citenamefont {Bellomo}\ \emph {et~al.}(2007)\citenamefont
  {Bellomo}, \citenamefont {{Lo Franco}},\ and\ \citenamefont
  {Compagno}}]{bellomo2007non}%
  \BibitemOpen
  \bibfield  {author} {\bibinfo {author} {\bibfnamefont {B.}~\bibnamefont
  {Bellomo}}, \bibinfo {author} {\bibfnamefont {R.}~\bibnamefont {{Lo
  Franco}}}, \ and\ \bibinfo {author} {\bibfnamefont {G.}~\bibnamefont
  {Compagno}},\ }\href@noop {} {\bibfield  {journal} {\bibinfo  {journal}
  {Phys. Rev. Lett.}\ }\textbf {\bibinfo {volume} {99}},\ \bibinfo {pages}
  {160502} (\bibinfo {year} {2007})}\BibitemShut {NoStop}%
\bibitem [{\citenamefont {{Lo Franco}}\ \emph {et~al.}(2014)\citenamefont {{Lo
  Franco}}, \citenamefont {D'Arrigo}, \citenamefont {Falci}, \citenamefont
  {Compagno},\ and\ \citenamefont {Paladino}}]{lo2014preserving}%
  \BibitemOpen
  \bibfield  {author} {\bibinfo {author} {\bibfnamefont {R.}~\bibnamefont {{Lo
  Franco}}}, \bibinfo {author} {\bibfnamefont {A.}~\bibnamefont {D'Arrigo}},
  \bibinfo {author} {\bibfnamefont {G.}~\bibnamefont {Falci}}, \bibinfo
  {author} {\bibfnamefont {G.}~\bibnamefont {Compagno}}, \ and\ \bibinfo
  {author} {\bibfnamefont {E.}~\bibnamefont {Paladino}},\ }\href@noop {}
  {\bibfield  {journal} {\bibinfo  {journal} {Phys. Rev. B}\ }\textbf {\bibinfo
  {volume} {90}},\ \bibinfo {pages} {054304} (\bibinfo {year}
  {2014})}\BibitemShut {NoStop}%
\bibitem [{\citenamefont {D'Arrigo}\ \emph {et~al.}(2014)\citenamefont
  {D'Arrigo}, \citenamefont {{Lo Franco}}, \citenamefont {Benenti},
  \citenamefont {Paladino},\ and\ \citenamefont {Falci}}]{d2014recovering}%
  \BibitemOpen
  \bibfield  {author} {\bibinfo {author} {\bibfnamefont {A.}~\bibnamefont
  {D'Arrigo}}, \bibinfo {author} {\bibfnamefont {R.}~\bibnamefont {{Lo
  Franco}}}, \bibinfo {author} {\bibfnamefont {G.}~\bibnamefont {Benenti}},
  \bibinfo {author} {\bibfnamefont {E.}~\bibnamefont {Paladino}}, \ and\
  \bibinfo {author} {\bibfnamefont {G.}~\bibnamefont {Falci}},\ }\href@noop {}
  {\bibfield  {journal} {\bibinfo  {journal} {Ann. Phys.}\ }\textbf {\bibinfo
  {volume} {350}},\ \bibinfo {pages} {211} (\bibinfo {year}
  {2014})}\BibitemShut {NoStop}%
\bibitem [{\citenamefont {Aolita}\ \emph {et~al.}(2015)\citenamefont {Aolita},
  \citenamefont {de~Melo},\ and\ \citenamefont {Davidovich}}]{aolita2015open}%
  \BibitemOpen
  \bibfield  {author} {\bibinfo {author} {\bibfnamefont {L.}~\bibnamefont
  {Aolita}}, \bibinfo {author} {\bibfnamefont {F.}~\bibnamefont {de~Melo}}, \
  and\ \bibinfo {author} {\bibfnamefont {L.}~\bibnamefont {Davidovich}},\
  }\href@noop {} {\bibfield  {journal} {\bibinfo  {journal} {Rep. Prog. Phys.}\
  }\textbf {\bibinfo {volume} {78}},\ \bibinfo {pages} {042001} (\bibinfo
  {year} {2015})}\BibitemShut {NoStop}%
\bibitem [{\citenamefont {Galve}\ \emph {et~al.}(2016)\citenamefont {Galve},
  \citenamefont {Zambrini},\ and\ \citenamefont {Maniscalco}}]{Galve2016}%
  \BibitemOpen
  \bibfield  {author} {\bibinfo {author} {\bibfnamefont {F.}~\bibnamefont
  {Galve}}, \bibinfo {author} {\bibfnamefont {R.}~\bibnamefont {Zambrini}}, \
  and\ \bibinfo {author} {\bibfnamefont {S.}~\bibnamefont {Maniscalco}},\
  }\href {http://dx.doi.org/10.1038/srep19607} {\bibfield  {journal} {\bibinfo
  {journal} {Sci. Rep.}\ }\textbf {\bibinfo {volume} {6}},\ \bibinfo {pages}
  {19607} (\bibinfo {year} {2016})}\BibitemShut {NoStop}%
\bibitem [{\citenamefont {Liu}\ \emph {et~al.}(2016)\citenamefont {Liu},
  \citenamefont {Hu}, \citenamefont {Huang}, \citenamefont {Li}, \citenamefont
  {Guo}, \citenamefont {Karlsson}, \citenamefont {Laine}, \citenamefont
  {Maniscalco}, \citenamefont {Macchiavello},\ and\ \citenamefont
  {Piilo}}]{Liu2016}%
  \BibitemOpen
  \bibfield  {author} {\bibinfo {author} {\bibfnamefont {B.-H.}\ \bibnamefont
  {Liu}}, \bibinfo {author} {\bibfnamefont {X.-M.}\ \bibnamefont {Hu}},
  \bibinfo {author} {\bibfnamefont {Y.-F.}\ \bibnamefont {Huang}}, \bibinfo
  {author} {\bibfnamefont {C.-F.}\ \bibnamefont {Li}}, \bibinfo {author}
  {\bibfnamefont {G.-C.}\ \bibnamefont {Guo}}, \bibinfo {author} {\bibfnamefont
  {A.}~\bibnamefont {Karlsson}}, \bibinfo {author} {\bibfnamefont {E.-M.}\
  \bibnamefont {Laine}}, \bibinfo {author} {\bibfnamefont {S.}~\bibnamefont
  {Maniscalco}}, \bibinfo {author} {\bibfnamefont {C.}~\bibnamefont
  {Macchiavello}}, \ and\ \bibinfo {author} {\bibfnamefont {J.}~\bibnamefont
  {Piilo}},\ }\href {http://dx.doi.org/10.1209/0295-5075/114/10005} {\bibfield
  {journal} {\bibinfo  {journal} {{EPL}}\ }\textbf {\bibinfo {volume} {114}},\
  \bibinfo {pages} {10005} (\bibinfo {year} {2016})}\BibitemShut {NoStop}%
\bibitem [{\citenamefont {Vasile}\ \emph {et~al.}(2011)\citenamefont {Vasile},
  \citenamefont {Olivares}, \citenamefont {Paris},\ and\ \citenamefont
  {Maniscalco}}]{Vasile2011}%
  \BibitemOpen
  \bibfield  {author} {\bibinfo {author} {\bibfnamefont {R.}~\bibnamefont
  {Vasile}}, \bibinfo {author} {\bibfnamefont {S.}~\bibnamefont {Olivares}},
  \bibinfo {author} {\bibfnamefont {M.~A.}\ \bibnamefont {Paris}}, \ and\
  \bibinfo {author} {\bibfnamefont {S.}~\bibnamefont {Maniscalco}},\ }\href
  {http://dx.doi.org/10.1103/PhysRevA.83.042321} {\bibfield  {journal}
  {\bibinfo  {journal} {Phys. Rev. A}\ }\textbf {\bibinfo {volume} {83}},\
  \bibinfo {pages} {042321} (\bibinfo {year} {2011})}\BibitemShut {NoStop}%
\bibitem [{\citenamefont {Schmidt}\ \emph {et~al.}(2011)\citenamefont
  {Schmidt}, \citenamefont {Negretti}, \citenamefont {Ankerhold}, \citenamefont
  {Calarco},\ and\ \citenamefont {Stockburger}}]{Schmidt2011}%
  \BibitemOpen
  \bibfield  {author} {\bibinfo {author} {\bibfnamefont {R.}~\bibnamefont
  {Schmidt}}, \bibinfo {author} {\bibfnamefont {A.}~\bibnamefont {Negretti}},
  \bibinfo {author} {\bibfnamefont {J.}~\bibnamefont {Ankerhold}}, \bibinfo
  {author} {\bibfnamefont {T.}~\bibnamefont {Calarco}}, \ and\ \bibinfo
  {author} {\bibfnamefont {J.~T.}\ \bibnamefont {Stockburger}},\ }\href
  {http://dx.doi.org/10.1103/PhysRevLett.107.130404} {\bibfield  {journal}
  {\bibinfo  {journal} {Phys. Rev. Lett.}\ }\textbf {\bibinfo {volume} {107}},\
  \bibinfo {pages} {130404} (\bibinfo {year} {2011})}\BibitemShut {NoStop}%
\bibitem [{\citenamefont {Bernardes}\ \emph {et~al.}(2016)\citenamefont
  {Bernardes}, \citenamefont {Peterson}, \citenamefont {Sarthour},
  \citenamefont {Souza}, \citenamefont {Monken}, \citenamefont {Roditi},
  \citenamefont {Oliveira},\ and\ \citenamefont {Santos}}]{Bernardes2016}%
  \BibitemOpen
  \bibfield  {author} {\bibinfo {author} {\bibfnamefont {N.~K.}\ \bibnamefont
  {Bernardes}}, \bibinfo {author} {\bibfnamefont {J.~P.~S.}\ \bibnamefont
  {Peterson}}, \bibinfo {author} {\bibfnamefont {R.~S.}\ \bibnamefont
  {Sarthour}}, \bibinfo {author} {\bibfnamefont {A.~M.}\ \bibnamefont {Souza}},
  \bibinfo {author} {\bibfnamefont {C.~H.}\ \bibnamefont {Monken}}, \bibinfo
  {author} {\bibfnamefont {I.}~\bibnamefont {Roditi}}, \bibinfo {author}
  {\bibfnamefont {I.~S.}\ \bibnamefont {Oliveira}}, \ and\ \bibinfo {author}
  {\bibfnamefont {M.~F.}\ \bibnamefont {Santos}},\ }\href
  {http://dx.doi.org/10.1038/srep33945} {\bibfield  {journal} {\bibinfo
  {journal} {Sci. Rep.}\ }\textbf {\bibinfo {volume} {6}},\ \bibinfo {pages}
  {33945} (\bibinfo {year} {2016})}\BibitemShut {NoStop}%
\bibitem [{\citenamefont {Souza}\ \emph {et~al.}(2013)\citenamefont {Souza},
  \citenamefont {Li}, \citenamefont {Soares-Pinto}, \citenamefont {Sarthour},
  \citenamefont {Oliveira}, \citenamefont {Huelga}, \citenamefont
  {Paternostro},\ and\ \citenamefont {Semi{\~a}o}}]{souza2013experimental}%
  \BibitemOpen
  \bibfield  {author} {\bibinfo {author} {\bibfnamefont {A.}~\bibnamefont
  {Souza}}, \bibinfo {author} {\bibfnamefont {J.}~\bibnamefont {Li}}, \bibinfo
  {author} {\bibfnamefont {D.}~\bibnamefont {Soares-Pinto}}, \bibinfo {author}
  {\bibfnamefont {R.}~\bibnamefont {Sarthour}}, \bibinfo {author}
  {\bibfnamefont {S.}~\bibnamefont {Oliveira}}, \bibinfo {author}
  {\bibfnamefont {S.}~\bibnamefont {Huelga}}, \bibinfo {author} {\bibfnamefont
  {M.}~\bibnamefont {Paternostro}}, \ and\ \bibinfo {author} {\bibfnamefont
  {F.}~\bibnamefont {Semi{\~a}o}},\ }\href@noop {} {\  (\bibinfo {year}
  {2013})},\ \Eprint {http://arxiv.org/abs/1308.5761} {arXiv:1308.5761
  [quant-ph]} \BibitemShut {NoStop}%
\bibitem [{\citenamefont {Orieux}\ \emph {et~al.}(2015)\citenamefont {Orieux},
  \citenamefont {DArrigo}, \citenamefont {Ferranti}, \citenamefont {{Lo
  Franco}}, \citenamefont {Benenti}, \citenamefont {Paladino}, \citenamefont
  {Falci}, \citenamefont {Sciarrino},\ and\ \citenamefont
  {Mataloni}}]{Orieux2015}%
  \BibitemOpen
  \bibfield  {author} {\bibinfo {author} {\bibfnamefont {A.}~\bibnamefont
  {Orieux}}, \bibinfo {author} {\bibfnamefont {A.}~\bibnamefont {DArrigo}},
  \bibinfo {author} {\bibfnamefont {G.}~\bibnamefont {Ferranti}}, \bibinfo
  {author} {\bibfnamefont {R.}~\bibnamefont {{Lo Franco}}}, \bibinfo {author}
  {\bibfnamefont {G.}~\bibnamefont {Benenti}}, \bibinfo {author} {\bibfnamefont
  {E.}~\bibnamefont {Paladino}}, \bibinfo {author} {\bibfnamefont
  {G.}~\bibnamefont {Falci}}, \bibinfo {author} {\bibfnamefont
  {F.}~\bibnamefont {Sciarrino}}, \ and\ \bibinfo {author} {\bibfnamefont
  {P.}~\bibnamefont {Mataloni}},\ }\href {http://dx.doi.org/10.1038/srep08575}
  {\bibfield  {journal} {\bibinfo  {journal} {Sci. Rep.}\ }\textbf {\bibinfo
  {volume} {5}},\ \bibinfo {pages} {8575} (\bibinfo {year} {2015})}\BibitemShut
  {NoStop}%
\bibitem [{\citenamefont {Liu}\ \emph {et~al.}(2011)\citenamefont {Liu},
  \citenamefont {Li}, \citenamefont {Huang}, \citenamefont {Li}, \citenamefont
  {Guo}, \citenamefont {Laine}, \citenamefont {Breuer},\ and\ \citenamefont
  {Piilo}}]{Liu2011}%
  \BibitemOpen
  \bibfield  {author} {\bibinfo {author} {\bibfnamefont {B.-H.}\ \bibnamefont
  {Liu}}, \bibinfo {author} {\bibfnamefont {L.}~\bibnamefont {Li}}, \bibinfo
  {author} {\bibfnamefont {Y.-F.}\ \bibnamefont {Huang}}, \bibinfo {author}
  {\bibfnamefont {C.-F.}\ \bibnamefont {Li}}, \bibinfo {author} {\bibfnamefont
  {G.-C.}\ \bibnamefont {Guo}}, \bibinfo {author} {\bibfnamefont {E.-M.}\
  \bibnamefont {Laine}}, \bibinfo {author} {\bibfnamefont {H.-P.}\ \bibnamefont
  {Breuer}}, \ and\ \bibinfo {author} {\bibfnamefont {J.}~\bibnamefont
  {Piilo}},\ }\href {http://dx.doi.org/10.1038/nphys2085} {\bibfield  {journal}
  {\bibinfo  {journal} {Nat. Phys.}\ }\textbf {\bibinfo {volume} {7}},\
  \bibinfo {pages} {931} (\bibinfo {year} {2011})}\BibitemShut {NoStop}%
\bibitem [{\citenamefont {Fanchini}\ \emph {et~al.}(2014)\citenamefont
  {Fanchini}, \citenamefont {Karpat}, \citenamefont {{\c{C}}akmak},
  \citenamefont {Castelano}, \citenamefont {Aguilar}, \citenamefont
  {Far{\'{\i}}as}, \citenamefont {Walborn}, \citenamefont {Ribeiro},\ and\
  \citenamefont {de~Oliveira}}]{Fanchini2014}%
  \BibitemOpen
  \bibfield  {author} {\bibinfo {author} {\bibfnamefont {F.~F.}\ \bibnamefont
  {Fanchini}}, \bibinfo {author} {\bibfnamefont {G.}~\bibnamefont {Karpat}},
  \bibinfo {author} {\bibfnamefont {B.}~\bibnamefont {{\c{C}}akmak}}, \bibinfo
  {author} {\bibfnamefont {L.~K.}\ \bibnamefont {Castelano}}, \bibinfo {author}
  {\bibfnamefont {G.~H.}\ \bibnamefont {Aguilar}}, \bibinfo {author}
  {\bibfnamefont {O.~J.}\ \bibnamefont {Far{\'{\i}}as}}, \bibinfo {author}
  {\bibfnamefont {S.~P.}\ \bibnamefont {Walborn}}, \bibinfo {author}
  {\bibfnamefont {P.~H.~S.}\ \bibnamefont {Ribeiro}}, \ and\ \bibinfo {author}
  {\bibfnamefont {M.~C.}\ \bibnamefont {de~Oliveira}},\ }\href
  {http://dx.doi.org/10.1103/PhysRevLett.112.210402} {\bibfield  {journal}
  {\bibinfo  {journal} {Phys. Rev. Lett.}\ }\textbf {\bibinfo {volume} {112}},\
  \bibinfo {pages} {210402} (\bibinfo {year} {2014})}\BibitemShut {NoStop}%
\bibitem [{\citenamefont {Xu}\ \emph {et~al.}(2013)\citenamefont {Xu},
  \citenamefont {Sun}, \citenamefont {Li}, \citenamefont {Xu}, \citenamefont
  {Guo}, \citenamefont {Andersson}, \citenamefont {{Lo Franco}},\ and\
  \citenamefont {Compagno}}]{Xu2013}%
  \BibitemOpen
  \bibfield  {author} {\bibinfo {author} {\bibfnamefont {J.-S.}\ \bibnamefont
  {Xu}}, \bibinfo {author} {\bibfnamefont {K.}~\bibnamefont {Sun}}, \bibinfo
  {author} {\bibfnamefont {C.-F.}\ \bibnamefont {Li}}, \bibinfo {author}
  {\bibfnamefont {X.-Y.}\ \bibnamefont {Xu}}, \bibinfo {author} {\bibfnamefont
  {G.-C.}\ \bibnamefont {Guo}}, \bibinfo {author} {\bibfnamefont
  {E.}~\bibnamefont {Andersson}}, \bibinfo {author} {\bibfnamefont
  {R.}~\bibnamefont {{Lo Franco}}}, \ and\ \bibinfo {author} {\bibfnamefont
  {G.}~\bibnamefont {Compagno}},\ }\href {http://dx.doi.org/10.1038/ncomms3851}
  {\bibfield  {journal} {\bibinfo  {journal} {Nat. Commun.}\ }\textbf {\bibinfo
  {volume} {4}},\ \bibinfo {pages} {2851} (\bibinfo {year} {2013})}\BibitemShut
  {NoStop}%
\bibitem [{\citenamefont {Xiong}\ \emph {et~al.}(2010)\citenamefont {Xiong},
  \citenamefont {Zhang}, \citenamefont {Wang},\ and\ \citenamefont
  {Wu}}]{Xiong2010}%
  \BibitemOpen
  \bibfield  {author} {\bibinfo {author} {\bibfnamefont {H.-N.}\ \bibnamefont
  {Xiong}}, \bibinfo {author} {\bibfnamefont {W.-M.}\ \bibnamefont {Zhang}},
  \bibinfo {author} {\bibfnamefont {X.}~\bibnamefont {Wang}}, \ and\ \bibinfo
  {author} {\bibfnamefont {M.-H.}\ \bibnamefont {Wu}},\ }\href
  {http://dx.doi.org/10.1103/PhysRevA.82.012105} {\bibfield  {journal}
  {\bibinfo  {journal} {Phys. Rev. A}\ }\textbf {\bibinfo {volume} {82}},\
  \bibinfo {pages} {012105} (\bibinfo {year} {2010})}\BibitemShut {NoStop}%
\bibitem [{\citenamefont {Batalh\~ao}\ \emph {et~al.}(2014)\citenamefont
  {Batalh\~ao}, \citenamefont {de~Moraes~Neto}, \citenamefont {de~Ponte},\ and\
  \citenamefont {Moussa}}]{Batalhao2014}%
  \BibitemOpen
  \bibfield  {author} {\bibinfo {author} {\bibfnamefont {T.~B.}\ \bibnamefont
  {Batalh\~ao}}, \bibinfo {author} {\bibfnamefont {G.~D.}\ \bibnamefont
  {de~Moraes~Neto}}, \bibinfo {author} {\bibfnamefont {M.~A.}\ \bibnamefont
  {de~Ponte}}, \ and\ \bibinfo {author} {\bibfnamefont {M.~H.~Y.}\ \bibnamefont
  {Moussa}},\ }\href {http://link.aps.org/doi/10.1103/PhysRevA.90.032105}
  {\bibfield  {journal} {\bibinfo  {journal} {Phys. Rev. A}\ }\textbf {\bibinfo
  {volume} {90}},\ \bibinfo {pages} {032105} (\bibinfo {year}
  {2014})}\BibitemShut {NoStop}%
\bibitem [{\citenamefont {Krovi}\ \emph {et~al.}(2007)\citenamefont {Krovi},
  \citenamefont {Oreshkov}, \citenamefont {Ryazanov},\ and\ \citenamefont
  {Lidar}}]{Krovi2007}%
  \BibitemOpen
  \bibfield  {author} {\bibinfo {author} {\bibfnamefont {H.}~\bibnamefont
  {Krovi}}, \bibinfo {author} {\bibfnamefont {O.}~\bibnamefont {Oreshkov}},
  \bibinfo {author} {\bibfnamefont {M.}~\bibnamefont {Ryazanov}}, \ and\
  \bibinfo {author} {\bibfnamefont {D.~A.}\ \bibnamefont {Lidar}},\ }\href
  {http://dx.doi.org/10.1103/PhysRevA.76.052117} {\bibfield  {journal}
  {\bibinfo  {journal} {Phys. Rev. A}\ }\textbf {\bibinfo {volume} {76}},\
  \bibinfo {pages} {052117} (\bibinfo {year} {2007})}\BibitemShut {NoStop}%
\bibitem [{\citenamefont {Chin}\ \emph {et~al.}(2013)\citenamefont {Chin},
  \citenamefont {Prior}, \citenamefont {Rosenbach}, \citenamefont
  {Caycedo-Soler}, \citenamefont {Huelga},\ and\ \citenamefont
  {Plenio}}]{Chin2013}%
  \BibitemOpen
  \bibfield  {author} {\bibinfo {author} {\bibfnamefont {A.~W.}\ \bibnamefont
  {Chin}}, \bibinfo {author} {\bibfnamefont {J.}~\bibnamefont {Prior}},
  \bibinfo {author} {\bibfnamefont {R.}~\bibnamefont {Rosenbach}}, \bibinfo
  {author} {\bibfnamefont {F.}~\bibnamefont {Caycedo-Soler}}, \bibinfo {author}
  {\bibfnamefont {S.~F.}\ \bibnamefont {Huelga}}, \ and\ \bibinfo {author}
  {\bibfnamefont {M.~B.}\ \bibnamefont {Plenio}},\ }\href
  {http://dx.doi.org/10.1038/nphys2515} {\bibfield  {journal} {\bibinfo
  {journal} {Nat. Phys.}\ }\textbf {\bibinfo {volume} {9}},\ \bibinfo {pages}
  {113} (\bibinfo {year} {2013})}\BibitemShut {NoStop}%
\bibitem [{\citenamefont {C{\'{a}}rdenas}\ \emph {et~al.}(2015)\citenamefont
  {C{\'{a}}rdenas}, \citenamefont {Paternostro},\ and\ \citenamefont
  {Semi{\~{a}}o}}]{Crdenas2015}%
  \BibitemOpen
  \bibfield  {author} {\bibinfo {author} {\bibfnamefont {P.~C.}\ \bibnamefont
  {C{\'{a}}rdenas}}, \bibinfo {author} {\bibfnamefont {M.}~\bibnamefont
  {Paternostro}}, \ and\ \bibinfo {author} {\bibfnamefont {F.~L.}\ \bibnamefont
  {Semi{\~{a}}o}},\ }\href {http://dx.doi.org/10.1103/PhysRevA.91.022122}
  {\bibfield  {journal} {\bibinfo  {journal} {Phys. Rev. A}\ }\textbf {\bibinfo
  {volume} {91}},\ \bibinfo {pages} {022122} (\bibinfo {year}
  {2015})}\BibitemShut {NoStop}%
\bibitem [{\citenamefont {Chen}\ \emph {et~al.}(2015)\citenamefont {Chen},
  \citenamefont {Lambert}, \citenamefont {Cheng}, \citenamefont {Chen},\ and\
  \citenamefont {Nori}}]{Chen2015}%
  \BibitemOpen
  \bibfield  {author} {\bibinfo {author} {\bibfnamefont {H.-B.}\ \bibnamefont
  {Chen}}, \bibinfo {author} {\bibfnamefont {N.}~\bibnamefont {Lambert}},
  \bibinfo {author} {\bibfnamefont {Y.-C.}\ \bibnamefont {Cheng}}, \bibinfo
  {author} {\bibfnamefont {Y.-N.}\ \bibnamefont {Chen}}, \ and\ \bibinfo
  {author} {\bibfnamefont {F.}~\bibnamefont {Nori}},\ }\href
  {http://dx.doi.org/10.1038/srep12753} {\bibfield  {journal} {\bibinfo
  {journal} {Sci. Rep.}\ }\textbf {\bibinfo {volume} {5}},\ \bibinfo {pages}
  {12753} (\bibinfo {year} {2015})}\BibitemShut {NoStop}%
\bibitem [{\citenamefont {Budini}\ \emph {et~al.}(1999)\citenamefont {Budini},
  \citenamefont {Chattah},\ and\ \citenamefont {C{\'{a}}ceres}}]{Budini1999}%
  \BibitemOpen
  \bibfield  {author} {\bibinfo {author} {\bibfnamefont {A.~A.}\ \bibnamefont
  {Budini}}, \bibinfo {author} {\bibfnamefont {A.~K.}\ \bibnamefont {Chattah}},
  \ and\ \bibinfo {author} {\bibfnamefont {M.~O.}\ \bibnamefont
  {C{\'{a}}ceres}},\ }\href {http://dx.doi.org/10.1088/0305-4470/32/4/007}
  {\bibfield  {journal} {\bibinfo  {journal} {J. Phys. A}\ }\textbf {\bibinfo
  {volume} {32}},\ \bibinfo {pages} {631} (\bibinfo {year} {1999})}\BibitemShut
  {NoStop}%
\bibitem [{\citenamefont {Budini}(2000)}]{Budini2000}%
  \BibitemOpen
  \bibfield  {author} {\bibinfo {author} {\bibfnamefont {A.~A.}\ \bibnamefont
  {Budini}},\ }\href {http://dx.doi.org/10.1103/PhysRevA.63.012106} {\bibfield
  {journal} {\bibinfo  {journal} {Phys. Rev. A}\ }\textbf {\bibinfo {volume}
  {63}},\ \bibinfo {pages} {012106} (\bibinfo {year} {2000})}\BibitemShut
  {NoStop}%
\bibitem [{\citenamefont {Budini}(2001)}]{Budini2001}%
  \BibitemOpen
  \bibfield  {author} {\bibinfo {author} {\bibfnamefont {A.~A.}\ \bibnamefont
  {Budini}},\ }\href {http://dx.doi.org/10.1103/PhysRevA.64.052110} {\bibfield
  {journal} {\bibinfo  {journal} {Phys. Rev. A}\ }\textbf {\bibinfo {volume}
  {64}},\ \bibinfo {pages} {052110} (\bibinfo {year} {2001})}\BibitemShut
  {NoStop}%
\bibitem [{\citenamefont {James}(1998)}]{James1998}%
  \BibitemOpen
  \bibfield  {author} {\bibinfo {author} {\bibfnamefont {D.~F.~V.}\
  \bibnamefont {James}},\ }\href {http://dx.doi.org/10.1103/PhysRevLett.81.317}
  {\bibfield  {journal} {\bibinfo  {journal} {Phys. Rev. Lett.}\ }\textbf
  {\bibinfo {volume} {81}},\ \bibinfo {pages} {317} (\bibinfo {year}
  {1998})}\BibitemShut {NoStop}%
\bibitem [{\citenamefont {Benedetti}\ \emph {et~al.}(2013)\citenamefont
  {Benedetti}, \citenamefont {Buscemi}, \citenamefont {Bordone},\ and\
  \citenamefont {Paris}}]{Benedetti2013}%
  \BibitemOpen
  \bibfield  {author} {\bibinfo {author} {\bibfnamefont {C.}~\bibnamefont
  {Benedetti}}, \bibinfo {author} {\bibfnamefont {F.}~\bibnamefont {Buscemi}},
  \bibinfo {author} {\bibfnamefont {P.}~\bibnamefont {Bordone}}, \ and\
  \bibinfo {author} {\bibfnamefont {M.~G.~A.}\ \bibnamefont {Paris}},\ }\href
  {http://dx.doi.org/10.1103/PhysRevA.87.052328} {\bibfield  {journal}
  {\bibinfo  {journal} {Phys. Rev. A}\ }\textbf {\bibinfo {volume} {87}},\
  \bibinfo {pages} {052328} (\bibinfo {year} {2013})}\BibitemShut {NoStop}%
\bibitem [{\citenamefont {Benedetti}\ and\ \citenamefont
  {Paris}(2014)}]{Benedetti2014}%
  \BibitemOpen
  \bibfield  {author} {\bibinfo {author} {\bibfnamefont {C.}~\bibnamefont
  {Benedetti}}\ and\ \bibinfo {author} {\bibfnamefont {M.~G.}\ \bibnamefont
  {Paris}},\ }\href {http://dx.doi.org/10.1016/j.physleta.2014.06.043}
  {\bibfield  {journal} {\bibinfo  {journal} {Phys. Lett. A}\ }\textbf
  {\bibinfo {volume} {378}},\ \bibinfo {pages} {2495} (\bibinfo {year}
  {2014})}\BibitemShut {NoStop}%
\bibitem [{\citenamefont {{Lo Franco}}\ and\ \citenamefont
  {Compagno}(2016)}]{lo2016overview}%
  \BibitemOpen
  \bibfield  {author} {\bibinfo {author} {\bibfnamefont {R.}~\bibnamefont {{Lo
  Franco}}}\ and\ \bibinfo {author} {\bibfnamefont {G.}~\bibnamefont
  {Compagno}},\ }\href@noop {} {\  (\bibinfo {year} {2016})},\ \Eprint
  {http://arxiv.org/abs/1608.05970} {arXiv:1608.05970 [quant-ph]} \BibitemShut
  {NoStop}%
\bibitem [{\citenamefont {Plenio}\ and\ \citenamefont
  {Knight}(1998)}]{Plenio1998}%
  \BibitemOpen
  \bibfield  {author} {\bibinfo {author} {\bibfnamefont {M.~B.}\ \bibnamefont
  {Plenio}}\ and\ \bibinfo {author} {\bibfnamefont {P.~L.}\ \bibnamefont
  {Knight}},\ }\href {http://dx.doi.org/10.1103/RevModPhys.70.101} {\bibfield
  {journal} {\bibinfo  {journal} {Rev. Mod. Phys.}\ }\textbf {\bibinfo {volume}
  {70}},\ \bibinfo {pages} {101} (\bibinfo {year} {1998})}\BibitemShut
  {NoStop}%
\bibitem [{\citenamefont {Dalibard}\ \emph {et~al.}(1992)\citenamefont
  {Dalibard}, \citenamefont {Castin},\ and\ \citenamefont
  {M{\o}lmer}}]{Dalibard1992}%
  \BibitemOpen
  \bibfield  {author} {\bibinfo {author} {\bibfnamefont {J.}~\bibnamefont
  {Dalibard}}, \bibinfo {author} {\bibfnamefont {Y.}~\bibnamefont {Castin}}, \
  and\ \bibinfo {author} {\bibfnamefont {K.}~\bibnamefont {M{\o}lmer}},\ }\href
  {http://dx.doi.org/10.1103/PhysRevLett.68.580} {\bibfield  {journal}
  {\bibinfo  {journal} {Phys. Rev. Lett.}\ }\textbf {\bibinfo {volume} {68}},\
  \bibinfo {pages} {580} (\bibinfo {year} {1992})}\BibitemShut {NoStop}%
\bibitem [{\citenamefont {Novikov}(1965)}]{NOVIKOV}%
  \BibitemOpen
  \bibfield  {author} {\bibinfo {author} {\bibfnamefont {E.~A.}\ \bibnamefont
  {Novikov}},\ }\href@noop {} {\bibfield  {journal} {\bibinfo  {journal} {Sov.
  Phys. JETP-URSS}\ }\textbf {\bibinfo {volume} {20}},\ \bibinfo {pages} {1290}
  (\bibinfo {year} {1965})}\BibitemShut {NoStop}%
\bibitem [{\citenamefont {Moss}\ and\ \citenamefont
  {McClintock}(1989)}]{moss1989noise}%
  \BibitemOpen
  \bibfield  {author} {\bibinfo {author} {\bibfnamefont {F.}~\bibnamefont
  {Moss}}\ and\ \bibinfo {author} {\bibfnamefont {P.}~\bibnamefont
  {McClintock}},\ }\href {https://books.google.com.br/books?id=Y7uaDQEACAAJ}
  {\emph {\bibinfo {title} {Noise in Nonlinear Dynamical Systems: Volume 1,
  Theory of Continuous Fokker-Planck Systems}}}\ (\bibinfo  {publisher}
  {Cambridge University Press},\ \bibinfo {year} {1989})\BibitemShut {NoStop}%
\bibitem [{\citenamefont {Vladimirov}(2002)}]{Vladimirov200208}%
  \BibitemOpen
  \bibfield  {author} {\bibinfo {author} {\bibfnamefont {V.~S.}\ \bibnamefont
  {Vladimirov}},\ }\href {http://amazon.com/o/ASIN/0415273560/} {\emph
  {\bibinfo {title} {Methods of the Theory of Generalized Functions (Analytical
  Methods and Special Functions)}}},\ \bibinfo {edition} {1st}\ ed.\ (\bibinfo
  {publisher} {CRC Press},\ \bibinfo {year} {2002})\BibitemShut {NoStop}%
\bibitem [{\citenamefont {Hall}\ \emph {et~al.}(2014)\citenamefont {Hall},
  \citenamefont {Cresser}, \citenamefont {Li},\ and\ \citenamefont
  {Andersson}}]{Hall2014}%
  \BibitemOpen
  \bibfield  {author} {\bibinfo {author} {\bibfnamefont {M.~J.~W.}\
  \bibnamefont {Hall}}, \bibinfo {author} {\bibfnamefont {J.~D.}\ \bibnamefont
  {Cresser}}, \bibinfo {author} {\bibfnamefont {L.}~\bibnamefont {Li}}, \ and\
  \bibinfo {author} {\bibfnamefont {E.}~\bibnamefont {Andersson}},\ }\href
  {http://dx.doi.org/10.1103/PhysRevA.89.042120} {\bibfield  {journal}
  {\bibinfo  {journal} {Phys. Rev. A}\ }\textbf {\bibinfo {volume} {89}},\
  \bibinfo {pages} {042120} (\bibinfo {year} {2014})}\BibitemShut {NoStop}%
\bibitem [{\citenamefont {Chru{\'{s}}ci{\'{n}}ski}\ and\ \citenamefont
  {Maniscalco}(2014)}]{Chruciski2014}%
  \BibitemOpen
  \bibfield  {author} {\bibinfo {author} {\bibfnamefont {D.}~\bibnamefont
  {Chru{\'{s}}ci{\'{n}}ski}}\ and\ \bibinfo {author} {\bibfnamefont
  {S.}~\bibnamefont {Maniscalco}},\ }\href
  {http://dx.doi.org/10.1103/PhysRevLett.112.120404} {\bibfield  {journal}
  {\bibinfo  {journal} {Phys. Rev. Lett.}\ }\textbf {\bibinfo {volume} {112}},\
  \bibinfo {pages} {120404} (\bibinfo {year} {2014})}\BibitemShut {NoStop}%
\bibitem [{\citenamefont {Lindblad}(1976)}]{Lindblad1976}%
  \BibitemOpen
  \bibfield  {author} {\bibinfo {author} {\bibfnamefont {G.}~\bibnamefont
  {Lindblad}},\ }\href {http://dx.doi.org/10.1007/BF01608499} {\bibfield
  {journal} {\bibinfo  {journal} {Commun. Math. Phys.}\ }\textbf {\bibinfo
  {volume} {48}},\ \bibinfo {pages} {119} (\bibinfo {year} {1976})}\BibitemShut
  {NoStop}%
\bibitem [{\citenamefont {Choi}(1975)}]{Choi1975}%
  \BibitemOpen
  \bibfield  {author} {\bibinfo {author} {\bibfnamefont {M.-D.}\ \bibnamefont
  {Choi}},\ }\href {http://dx.doi.org/10.1016/0024-3795(75)90075-0} {\bibfield
  {journal} {\bibinfo  {journal} {Linear Algebra Appl.}\ }\textbf {\bibinfo
  {volume} {10}},\ \bibinfo {pages} {285} (\bibinfo {year} {1975})}\BibitemShut
  {NoStop}%
\bibitem [{\citenamefont {Jamio{\l}kowski}(1972)}]{Jamiokowski1972}%
  \BibitemOpen
  \bibfield  {author} {\bibinfo {author} {\bibfnamefont {A.}~\bibnamefont
  {Jamio{\l}kowski}},\ }\href {http://dx.doi.org/10.1016/0034-4877(72)90011-0}
  {\bibfield  {journal} {\bibinfo  {journal} {Rep. Math. Phys.}\ }\textbf
  {\bibinfo {volume} {3}},\ \bibinfo {pages} {275} (\bibinfo {year}
  {1972})}\BibitemShut {NoStop}%
\bibitem [{\citenamefont {Nielsen}\ and\ \citenamefont
  {Chuang}(2012)}]{NielsenChuang201206}%
  \BibitemOpen
  \bibfield  {author} {\bibinfo {author} {\bibfnamefont {M.~A.}\ \bibnamefont
  {Nielsen}}\ and\ \bibinfo {author} {\bibfnamefont {I.~L.}\ \bibnamefont
  {Chuang}},\ }\href {http://amazon.com/o/ASIN/0511976666/} {\emph {\bibinfo
  {title} {Quantum Computation and Quantum Information: 10th Anniversary
  Edition}}}\ (\bibinfo  {publisher} {Cambridge University Press},\ \bibinfo
  {year} {2012})\BibitemShut {NoStop}%
\bibitem [{\citenamefont {Lorenzo}\ \emph {et~al.}(2013)\citenamefont
  {Lorenzo}, \citenamefont {Plastina},\ and\ \citenamefont
  {Paternostro}}]{Lorenzo2013}%
  \BibitemOpen
  \bibfield  {author} {\bibinfo {author} {\bibfnamefont {S.}~\bibnamefont
  {Lorenzo}}, \bibinfo {author} {\bibfnamefont {F.}~\bibnamefont {Plastina}}, \
  and\ \bibinfo {author} {\bibfnamefont {M.}~\bibnamefont {Paternostro}},\
  }\href {http://dx.doi.org/10.1103/PhysRevA.88.020102} {\bibfield  {journal}
  {\bibinfo  {journal} {Phys. Rev. A}\ }\textbf {\bibinfo {volume} {88}},\
  \bibinfo {pages} {020102} (\bibinfo {year} {2013})}\BibitemShut {NoStop}%
\bibitem [{\citenamefont {Ballentine}(1998)}]{Ballentine199805}%
  \BibitemOpen
  \bibfield  {author} {\bibinfo {author} {\bibfnamefont {L.~E.}\ \bibnamefont
  {Ballentine}},\ }\href {http://amazon.com/o/ASIN/9810227078/} {\emph
  {\bibinfo {title} {Quantum Mechanics: A Modern Development}}},\ \bibinfo
  {edition} {2nd}\ ed.\ (\bibinfo  {publisher} {Wspc},\ \bibinfo {year}
  {1998})\BibitemShut {NoStop}%
\bibitem [{\citenamefont {Lidar}\ \emph {et~al.}(2001)\citenamefont {Lidar},
  \citenamefont {Bihary},\ and\ \citenamefont {Whaley}}]{Lidar2001}%
  \BibitemOpen
  \bibfield  {author} {\bibinfo {author} {\bibfnamefont {D.~A.}\ \bibnamefont
  {Lidar}}, \bibinfo {author} {\bibfnamefont {Z.}~\bibnamefont {Bihary}}, \
  and\ \bibinfo {author} {\bibfnamefont {K.}~\bibnamefont {Whaley}},\ }\href
  {http://dx.doi.org/10.1016/S0301-0104(01)00330-5} {\bibfield  {journal}
  {\bibinfo  {journal} {Chem. Phys.}\ }\textbf {\bibinfo {volume} {268}},\
  \bibinfo {pages} {35} (\bibinfo {year} {2001})}\BibitemShut {NoStop}%
\bibitem [{\citenamefont {Moussa}\ \emph {et~al.}(1996)\citenamefont {Moussa},
  \citenamefont {Mizrahi},\ and\ \citenamefont {Caldeira}}]{Moussa1996}%
  \BibitemOpen
  \bibfield  {author} {\bibinfo {author} {\bibfnamefont {M.}~\bibnamefont
  {Moussa}}, \bibinfo {author} {\bibfnamefont {S.}~\bibnamefont {Mizrahi}}, \
  and\ \bibinfo {author} {\bibfnamefont {A.}~\bibnamefont {Caldeira}},\ }\href
  {http://dx.doi.org/10.1016/0375-9601(96)00577-4} {\bibfield  {journal}
  {\bibinfo  {journal} {Phys. Lett. A}\ }\textbf {\bibinfo {volume} {221}},\
  \bibinfo {pages} {145} (\bibinfo {year} {1996})}\BibitemShut {NoStop}%
\bibitem [{\citenamefont {Poyatos}\ \emph {et~al.}(1996)\citenamefont
  {Poyatos}, \citenamefont {Cirac},\ and\ \citenamefont
  {Zoller}}]{QuantReservEngPoyatos1996}%
  \BibitemOpen
  \bibfield  {author} {\bibinfo {author} {\bibfnamefont {J.~F.}\ \bibnamefont
  {Poyatos}}, \bibinfo {author} {\bibfnamefont {J.~I.}\ \bibnamefont {Cirac}},
  \ and\ \bibinfo {author} {\bibfnamefont {P.}~\bibnamefont {Zoller}},\ }\href
  {\doibase 10.1103/PhysRevLett.77.4728} {\bibfield  {journal} {\bibinfo
  {journal} {Phys. Rev. Lett.}\ }\textbf {\bibinfo {volume} {77}},\ \bibinfo
  {pages} {4728} (\bibinfo {year} {1996})}\BibitemShut {NoStop}%
\bibitem [{\citenamefont {Turchette}\ \emph {et~al.}(2000)\citenamefont
  {Turchette}, \citenamefont {Myatt}, \citenamefont {King}, \citenamefont
  {Sackett}, \citenamefont {Kielpinski}, \citenamefont {Itano}, \citenamefont
  {Monroe},\ and\ \citenamefont {Wineland}}]{DecoherenceMyatt200}%
  \BibitemOpen
  \bibfield  {author} {\bibinfo {author} {\bibfnamefont {Q.~A.}\ \bibnamefont
  {Turchette}}, \bibinfo {author} {\bibfnamefont {C.~J.}\ \bibnamefont
  {Myatt}}, \bibinfo {author} {\bibfnamefont {B.~E.}\ \bibnamefont {King}},
  \bibinfo {author} {\bibfnamefont {C.~A.}\ \bibnamefont {Sackett}}, \bibinfo
  {author} {\bibfnamefont {D.}~\bibnamefont {Kielpinski}}, \bibinfo {author}
  {\bibfnamefont {W.~M.}\ \bibnamefont {Itano}}, \bibinfo {author}
  {\bibfnamefont {C.}~\bibnamefont {Monroe}}, \ and\ \bibinfo {author}
  {\bibfnamefont {D.~J.}\ \bibnamefont {Wineland}},\ }\href {\doibase
  10.1103/PhysRevA.62.053807} {\bibfield  {journal} {\bibinfo  {journal} {Phys.
  Rev. A}\ }\textbf {\bibinfo {volume} {62}},\ \bibinfo {pages} {053807}
  (\bibinfo {year} {2000})}\BibitemShut {NoStop}%
\bibitem [{\citenamefont {Kienzler}\ \emph {et~al.}(2014)\citenamefont
  {Kienzler}, \citenamefont {Lo}, \citenamefont {Keitch}, \citenamefont
  {de~Clercq}, \citenamefont {Leupold}, \citenamefont {Lindenfelser},
  \citenamefont {Marinelli}, \citenamefont {Negnevitsky},\ and\ \citenamefont
  {Home}}]{QuantHarmoReservEngKienzler2014}%
  \BibitemOpen
  \bibfield  {author} {\bibinfo {author} {\bibfnamefont {D.}~\bibnamefont
  {Kienzler}}, \bibinfo {author} {\bibfnamefont {H.-Y.}\ \bibnamefont {Lo}},
  \bibinfo {author} {\bibfnamefont {B.}~\bibnamefont {Keitch}}, \bibinfo
  {author} {\bibfnamefont {L.}~\bibnamefont {de~Clercq}}, \bibinfo {author}
  {\bibfnamefont {F.}~\bibnamefont {Leupold}}, \bibinfo {author} {\bibfnamefont
  {F.}~\bibnamefont {Lindenfelser}}, \bibinfo {author} {\bibfnamefont
  {M.}~\bibnamefont {Marinelli}}, \bibinfo {author} {\bibfnamefont
  {V.}~\bibnamefont {Negnevitsky}}, \ and\ \bibinfo {author} {\bibfnamefont
  {J.~P.}\ \bibnamefont {Home}},\ }\href
  {http://dx.doi.org/10.1126/science.1261033} {\bibfield  {journal} {\bibinfo
  {journal} {Science}\ } (\bibinfo {year} {2014})}\BibitemShut {NoStop}%
\bibitem [{\citenamefont {Leibfried}\ \emph {et~al.}(2003)\citenamefont
  {Leibfried}, \citenamefont {Blatt}, \citenamefont {Monroe},\ and\
  \citenamefont {Wineland}}]{QuantumDynamicsLeibfried2003}%
  \BibitemOpen
  \bibfield  {author} {\bibinfo {author} {\bibfnamefont {D.}~\bibnamefont
  {Leibfried}}, \bibinfo {author} {\bibfnamefont {R.}~\bibnamefont {Blatt}},
  \bibinfo {author} {\bibfnamefont {C.}~\bibnamefont {Monroe}}, \ and\ \bibinfo
  {author} {\bibfnamefont {D.}~\bibnamefont {Wineland}},\ }\href {\doibase
  10.1103/RevModPhys.75.281} {\bibfield  {journal} {\bibinfo  {journal} {Rev.
  Mod. Phys.}\ }\textbf {\bibinfo {volume} {75}},\ \bibinfo {pages} {281}
  (\bibinfo {year} {2003})}\BibitemShut {NoStop}%
\bibitem [{\citenamefont {Roos}\ \emph {et~al.}(2004)\citenamefont {Roos},
  \citenamefont {Lancaster}, \citenamefont {Riebe}, \citenamefont {H\"affner},
  \citenamefont {H\"ansel}, \citenamefont {Gulde}, \citenamefont {Becher},
  \citenamefont {Eschner}, \citenamefont {Schmidt-Kaler},\ and\ \citenamefont
  {Blatt}}]{QuantumTomographyRoss2004}%
  \BibitemOpen
  \bibfield  {author} {\bibinfo {author} {\bibfnamefont {C.~F.}\ \bibnamefont
  {Roos}}, \bibinfo {author} {\bibfnamefont {G.~P.~T.}\ \bibnamefont
  {Lancaster}}, \bibinfo {author} {\bibfnamefont {M.}~\bibnamefont {Riebe}},
  \bibinfo {author} {\bibfnamefont {H.}~\bibnamefont {H\"affner}}, \bibinfo
  {author} {\bibfnamefont {W.}~\bibnamefont {H\"ansel}}, \bibinfo {author}
  {\bibfnamefont {S.}~\bibnamefont {Gulde}}, \bibinfo {author} {\bibfnamefont
  {C.}~\bibnamefont {Becher}}, \bibinfo {author} {\bibfnamefont
  {J.}~\bibnamefont {Eschner}}, \bibinfo {author} {\bibfnamefont
  {F.}~\bibnamefont {Schmidt-Kaler}}, \ and\ \bibinfo {author} {\bibfnamefont
  {R.}~\bibnamefont {Blatt}},\ }\href {\doibase 10.1103/PhysRevLett.92.220402}
  {\bibfield  {journal} {\bibinfo  {journal} {Phys. Rev. Lett.}\ }\textbf
  {\bibinfo {volume} {92}},\ \bibinfo {pages} {220402} (\bibinfo {year}
  {2004})}\BibitemShut {NoStop}%
\bibitem [{\citenamefont {Palao}\ \emph {et~al.}(2001)\citenamefont {Palao},
  \citenamefont {Kosloff},\ and\ \citenamefont {Gordon}}]{Palao2001}%
  \BibitemOpen
  \bibfield  {author} {\bibinfo {author} {\bibfnamefont {J.~P.}\ \bibnamefont
  {Palao}}, \bibinfo {author} {\bibfnamefont {R.}~\bibnamefont {Kosloff}}, \
  and\ \bibinfo {author} {\bibfnamefont {J.~M.}\ \bibnamefont {Gordon}},\
  }\href {http://dx.doi.org/10.1103/PhysRevE.64.056130} {\bibfield  {journal}
  {\bibinfo  {journal} {Phys. Rev. E}\ }\textbf {\bibinfo {volume} {64}},\
  \bibinfo {pages} {056130} (\bibinfo {year} {2001})}\BibitemShut {NoStop}%
\bibitem [{\citenamefont {Rezek}\ \emph {et~al.}(2009)\citenamefont {Rezek},
  \citenamefont {Salamon}, \citenamefont {Hoffmann},\ and\ \citenamefont
  {Kosloff}}]{Rezek2009}%
  \BibitemOpen
  \bibfield  {author} {\bibinfo {author} {\bibfnamefont {Y.}~\bibnamefont
  {Rezek}}, \bibinfo {author} {\bibfnamefont {P.}~\bibnamefont {Salamon}},
  \bibinfo {author} {\bibfnamefont {K.~H.}\ \bibnamefont {Hoffmann}}, \ and\
  \bibinfo {author} {\bibfnamefont {R.}~\bibnamefont {Kosloff}},\ }\href
  {http://dx.doi.org/10.1209/0295-5075/85/30008} {\bibfield  {journal}
  {\bibinfo  {journal} {{EPL}}\ }\textbf {\bibinfo {volume} {85}},\ \bibinfo
  {pages} {30008} (\bibinfo {year} {2009})}\BibitemShut {NoStop}%
\end{thebibliography}
\end{document}